\documentclass[letter]{aa}  

\usepackage{graphicx}
\usepackage{txfonts}
\usepackage{lipsum}
\usepackage{subcaption}         % necessary for continued figures, example in section 3
\usepackage{lscape}             % to rotate a single page table, example in appendix.
\usepackage{placeins}           % useful with \FloatBarrier, to keep 
\usepackage{xcolor}

\usepackage[normalem]{ulem}

\newcommand{\mesa}{\texttt{MESA}}

\newcommand{\approxai}{\texttt{Approx}\_\texttt{A56}}
\newcommand{\approxaii}{\texttt{Approx}\_\texttt{A64}}
\newcommand{\bxlone}{\texttt{BXL1}}
\newcommand{\middmdot}{$\dot{M}_{\mathrm{moderate}~}$}
\newcommand{\highmdot}{$\dot{M}_{\mathrm{high}~}$}
\newcommand{\hydros}{$^{1}\mathrm{H}$}

\newcommand{\carbos}{$^{12}\mathrm{C}$}
\newcommand{\neos}{$^{20}\mathrm{Ne}$}

\newcommand{\silicos}{$^{28}\mathrm{Si}$}

\newcommand{\iroh}{$^{56}\mathrm{Fe}$}

\begin{document}

%%%%%%%%%%%%%%%%%%%%%%%%%%%%%%%%%%%%%%%%
% if you use custom commands in your title,
% ensure to check your title when submitting!
%%%%%%%%%%%%%%%%%%%%%%%%%%%%%%%%%%%%%%%%
   \title{On the superbursts as photodisintegrators of ashes from the rp-process}

   \subtitle{- Numerical simulations of superbursts with MESA II-}

%%%%%%%%%%%%%%%%%%%%%%%%%%%%%%%%%%%%%%%%
% Please separate each author with the \and command
%
% Please do not include ORCIDs next to author names.
% Only ORCIDs authenticated by individual authors in EDPS
% editorial system will be taken into account.
% ORCIDs included here will be removed.
%%%%%%%%%%%%%%%%%%%%%%%%%%%%%%%%%%%%%%%%

   \author{M. Nava-Callejas\inst{1}\fnmsep\thanks{Corresponding author: martin.javier.nava.callejas@ulb.be}
        \and S. Goriely,\inst{1}
        \and N. Chamel,\inst{1}
        }

   \institute{Institut d'Astronomie et d'Astrophysique, Université Libre de Bruxelles, 1050 Bruxelles, Belgique}

   \date{Received September 30, 20XX}

% \abstract{}{}{}{}{}
% 5 {} token are mandatory 
  \abstract
  % context
   {Superbursts are very energetic explosions in the crust of neutron stars in Low-Mass X-ray Binaries (LMXBs). Given that the temperatures reached during the explosion are well-above 4 GK and the ashes of the rp-process being the fuel behind superbursts, these events have been proposed as sites where photodisintegration of heavy elements play a larger role than the \carbos{} fusion.}
  % aims
   {The objective of the present study is to  explore in details the impact of photodisintegrations of rp-process ashes on the properties of superbursts and accordingly their effects on the initial \carbos{} abundance needed to trigger an explosion.} 
  % methods
   {For simulating superbursts, we employ the public code \mesa\ v.24.08.1 which is well suited to simulated energetic explosions at the surface of the accreting neutron star. As an alternative to the canonical fuel composition made of $20\%$ of \carbos{} and $80\%$ of $^{56}$Fe, we consider the accretion of material with different enrichments in \carbos{} complemented with a scaled distribution of ashes from the rp-process.
   }
  % results
   {We find that superbursts can be triggered with a \carbos{} mass fraction of no more than $10\%$, i.e. significantly lower than the value of 20\% usually needed, as long as one considers for the remaining $90\%$ a scaled distribution of ashes from the rp-process. The luminosity of these superbursts is comparable to that obtained by burning $20\%$ of \carbos{} and $80\%$ of $^{56}$Fe. In contrast, using $10\%$ of \carbos{} with $90\%$ of $^{56}$Fe produces a less luminous superbursts. We also find that $^{56}$Fe remains the most abundant nuclide in the ashes of superbursts. Additionally, we show that although photodisintegrations play a key role in the nucleosynthesis, the neutron, proton and alpha recaptures are actually responsible for releasing energy at late times during the explosion. Finally, we also identify that the presence of $A = 64$ nuclides or $^{20}$Ne is important to alleviate the required amount of \carbos{} needed to trigger the superbust ignition.}
%conclusions
   {The detailed composition of the accreted material, and not solely its \carbos{} content, substantially impact the superburst ignition, luminosity and nucleosynthesis.}

   \maketitle
   \nolinenumbers

%%%%%%%%%%%%%%%%%%%%%%%%%%%%%%%%%%%%%%%%%%%%%%%%%%%%%%%%%%%%%%%%%%%%%%%%%%%%%%%%%%%%%%%%%%%%%%%%%%%%%%%%%%%%%%%%%%%%%%%%%%%%%%%%%%%%%%%%%%%%%%%%%%%%%%%%%%%%%%%%%%%%%%%%%%%%%%%%%%%%%%%%%%%%%%%%%%%%%%%%%%%%%%%%%%%%%%%%%%%%%%%%%%%
%%%%%%%%%%%%%%%%%%%%%%%%%%%%%%%%%%%%%%%%%%%%%%%%%%%%%%%%%%%%%%%%%%%%%%%%%%%%%%%%%%%%%%%%%%%%%%%%%%%%%%%%%%%%%%%%%%%%%%%%%%%%%%%%%%%%%%%%%%%%%%%%%%%%%%%%%%%%%%%%%%%%%%%%%%%%%%%%%%%%%%%%%%%%%%%%%%%%%%%%%%%%%%%%%%%%%%%%%%%%%%%%%%%
\section{Introduction}
%%%%%%%%%%%%%%%%%%%%%%%%%%%%%%%%%%%%%%%%%%%%%%%%%%%%%%%%%%%%%%%%%%%%%%%%%%%%%%%%%%%%%%%%%%%%%%%%%%%%%%%%%%%%%%%%%%%%%%%%%%%%%%%%%%%%%%%%%%%%%%%%%%%%%%%%%%%%%%%%%%%%%%%%%%%%%%%%%%%%%%%%%%%%%%%%%%%%%%%%%%%%%%%%%%%%%%%%%%%%%%%%%%%

To model superbursts - thermonuclear explosions over the surface of neutron stars, driven by unstable \carbos{} burning \citep{Wijnands_2001, Cumming_2001, Strohmayer_2002a, Kuulkers_2004a} - it has been argued that a \carbos{} mass fraction of $20\%$ is required not only to trigger but also to power these highly energetic events. In this work, we show
that a \carbos{} mass fraction as low as $10\%$ is sufficient to power a superburst, provided that the remaining $90\%$ consists of a distribution of rp-process ashes rather than of a single fiducial nuclide (e.g. $80\%$ of \iroh), as commonly assumed.
The $20\%$ threshold can be traced back to the numerical simulations of \citet{Cumming_2004, Cumming_2006}. In these studies, superburst lightcurves were fitted using two parameters: a specific energy $E_\mathrm{super}$, and a column depth, $y_\mathrm{ignition}$, at which this energy was deposited. However, the thermonuclear runaway was not simulated but only the cooling phase, arguing that the injection of $E_\mathrm{super}$ at $y_\mathrm{ignition}$ sufficed to reproduce the observed cooling. The limitations of this two-parameter description might already be perceived in the multi-zone superburst simulations by \citet{KH2011}, who found that the ignition column depth for superbursts was systematically lower by almost an order of magnitude with respect to the ones reported by \citet{Cumming_2004, Cumming_2006}. Regarding $E_\mathrm{super}$, \citet{Schatz_2003} suggested that photodisintegration reactions could contribute significantly more than carbon fusion. Unlike \citet{Cumming_2004, Cumming_2006}, \citet{Schatz_2003} performed a numerical calculation that used as initial conditions a mass distribution which resembled the ashes produced by the rp-process \citep{Schatz_2001, Meisel_2018}. However, these calculations suffered from several limitations (see the discussion in \citet{NC_2026A&A}, hereafter NC26). In particular, their one-zone model (OZM) is not capable of fully resolving the  heating and cooling phases of the superburst.
Since the rp-process ashes constitute the actual fuel for superbursts, they should be considered as initial conditions for multi-zone simulations of superbursts. 
However, implementing them as initial conditions for superbursts has not been addressed so far due to the computational challenge that it implies, as it requires to follow the coupled evolution of nuclear burning , energy transport, and hydrodynamics while simultaneously resolving a broad range of nuclear species \citep{KH2011, KH2012}. In this work, we take a first step in addressing this challenge by using multi-zone simulations of superbursts.

%%%%%%%%%%%%%%%%%%%%%%%%%%%%%%%%%%%%%%%%%%%%%%%%%%%%%%%%%%%%%%%%%%%%%%%%%%%%%%%%%%%%%%%%%%%%%%%%%%%%%%%%%%%%%%%%%%%%%%%%%%%%%%%%%%%%%%%%%%%%%%%%%%%%%%%%%%%%%%%%%%%%%%%%%%%%%%%%%%%%%%%%%%%%%%%%%%%%%%%%%%%%%%%%%%%%%%%%%%%%%%%%%%%
%%%%%%%%%%%%%%%%%%%%%%%%%%%%%%%%%%%%%%%%%%%%%%%%%%%%%%%%%%%%%%%%%%%%%%%%%%%%%%%%%%%%%%%%%%%%%%%%%%%%%%%%%%%%%%%%%%%%%%%%%%%%%%%%%%%%%%%%%%%%%%%%%%%%%%%%%%%%%%%%%%%%%%%%%%%%%%%%%%%%%%%%%%%%%%%%%%%%%%%%%%%%%%%%%%%%%%%%%%%%%%%%%%%

%%%%%%%%%%%%%%%%%%%%%%%%%%%%%%%%%%%%%%%%%%%%%%%%%%%%%%%%%%%%%%%%%%%%%%%%%%%%%%%%%%%%%%%%%%%%%%%%%%%%%%%%%%%%%%%%%%%%%%%%%%%%%%%%%%%%%%%%%%%%%%%%%%%%%%%%%%%%%%%%%%%%%%%%%%%%%%%%%%%%%%%%%%%%%%%%%%%%%%%%%%%%%%%%%%%%%%%%%%%%%%%%%%%
%%%%%%%%%%%%%%%%%%%%%%%%%%%%%%%%%%%%%%%%%%%%%%%%%%%%%%%%%%%%%%%%%%%%%%%%%%%%%%%%%%%%%%%%%%%%%%%%%%%%%%%%%%%%%%%%%%%%%%%%%%%%%%%%%%%%%%%%%%%%%%%%%%%%%%%%%%%%%%%%%%%%%%%%%%%%%%%%%%%%%%%%%%%%%%%%%%%%%%%%%%%%%%%%%%%%%%%%%%%%%%%%%%%
\section{Methodology}\label{section:methods}
%%%%%%%%%%%%%%%%%%%%%%%%%%%%%%%%%%%%%%%%%%%%%%%%%%%%%%%%%%%%%%%%%%%%%%%%%%%%%%%%%%%%%%%%%%%%%%%%%%%%%%%%%%%%%%%%%%%%%%%%%%%%%%%%%%%%%%%%%%%%%%%%%%%%%%%%%%%%%%%%%%%%%%%%%%%%%%%%%%%%%%%%%%%%%%%%%%%%%%%%%%%%%%%%%%%%%%%%%%%%%%%%%%%

% - - - - - - - - - - - - - - - - - - - - - - - - - - - - - - - - - - - - - - - - - - - - - - - - - - - - - %
% 1. MESA
We perform our numerical simulations using the public code \mesa{} v.24.08.1 \citep{Jermyn_2023aa, Paxton_2019aa, Paxton_2018aa, Paxton_2015aa, Paxton_2013aa, Paxton_2011aa}. Our inlists are based on those used in NC26. We summarize below the parameters relevant to the present simulations.
% - - - - - - - - - - - - - - - - - - - - - - - - - - - - - - - - - - - - - - - - - - - - - - - - - - - - - %

% - - - - - - - - - - - - - - - - - - - - - - - - - - - - - - - - - - - - - - - - - - - - - - - - - - - - - %
% 2. Simulation setup
We consider two reaction networks. The first, \approxai, was introduced in NC26 and includes 190 nuclides in and above the valley of stability from $A=1$ to $A=56$. The second, \texttt{BXL1}, contains 636 nuclides from $A=1$ to $A=120$. We adopt the same mass accretion rates and base luminosities as in NC26: $\dot{M}_\mathrm{moderate}=5.88\times10^{-9}$ $M_{\odot}~\mathrm{yr}^{-1}$ ($\approx0.3\dot{M}_\mathrm{Edd}$) with $L_b=25L_{\odot}$, and $\dot{M}_\mathrm{high}=2\times10^{-8}$ $M_{\odot}~\mathrm{yr}^{-1}$ ($\approx\dot{M}_\mathrm{Edd}$) with $L_b=50L_{\odot}$. These base luminosities correspond to 0.26 and 0.16 MeV per baryon, respectively. Due to current uncertainties of the carbon-fusion rate \citep[see, e.g.,][]{Tumino18, Monpribat22, NC_2026A&A}, we adopt the standard CF88 rate \citep{CF88_bib}. 
We perform three sets of simulations with different accreted compositions. The first uses the \approxai{} network with initial mass fractions $X(^{12}\mathrm{C})=0.20$ and $X(^{56}\mathrm{Fe})=0.80$, corresponding to the canonical superburst composition \citep{Cumming_2006,KH2011}, hereafter Scenario A. The second uses the same network with $X(^{12}\mathrm{C})=0.10$ and $X(^{56}\mathrm{Fe})=0.90$, Scenario B. The third, Scenario C, uses the \texttt{BXL1} network with $X(^{12}\mathrm{C})=0.10$, while the remaining $90\%$ is given by a scaled distribution of ashes from the rp-process. This distribution was obtained from a separate calculation at $\dot{M}=3\times10^{-9}$ $M_{\odot}~\mathrm{yr}^{-1}$. Further details of this calculation are provided in Appendix~\ref{appendix:rp_ashes_calcs}.
For a brief discussion on the sensitivity of our superburst lightcurves and distribution of ashes to the size of the network, we refer the reader to \ref{appendix:network_size}.
% - - - - - - - - - - - - - - - - - - - - - - - - - - - - - - - - - - - - - - - - - - - - - - - - - - - - - %

%%%%%%%%%%%%%%%%%%%%%%%%%%%%%%%%%%%%%%%%%%%%%%%%%%%%%%%%%%%%%%%%%%%%%%%%%%%%%%%%%%%%%%%%%%%%%%%%%%%%%%%%%%%%%%%%%%%%%%%%%%%%%%%%%%%%%%%%%%%%%%%%%%%%%%%%%%%%%%%%%%%%%%%%%%%%%%%%%%%%%%%%%%%%%%%%%%%%%%%%%%%%%%%%%%%%%%%%%%%%%%%%%%%
%%%%%%%%%%%%%%%%%%%%%%%%%%%%%%%%%%%%%%%%%%%%%%%%%%%%%%%%%%%%%%%%%%%%%%%%%%%%%%%%%%%%%%%%%%%%%%%%%%%%%%%%%%%%%%%%%%%%%%%%%%%%%%%%%%%%%%%%%%%%%%%%%%%%%%%%%%%%%%%%%%%%%%%%%%%%%%%%%%%%%%%%%%%%%%%%%%%%%%%%%%%%%%%%%%%%%%%%%%%%%%%%%%%

%%%%%%%%%%%%%%%%%%%%%%%%%%%%%%%%%%%%%%%%%%%%%%%%%%%%%%%%%%%%%%%%%%%%%%%%%%%%%%%%%%%%%%%%%%%%%%%%%%%%%%%%%%%%%%%%%%%%%%%%%%%%%%%%%%%%%%%%%%%%%%%%%%%%%%%%%%%%%%%%%%%%%%%%%%%%%%%%%%%%%%%%%%%%%%%%%%%%%%%%%%%%%%%%%%%%%%%%%%%%%%%%%%%
%%%%%%%%%%%%%%%%%%%%%%%%%%%%%%%%%%%%%%%%%%%%%%%%%%%%%%%%%%%%%%%%%%%%%%%%%%%%%%%%%%%%%%%%%%%%%%%%%%%%%%%%%%%%%%%%%%%%%%%%%%%%%%%%%%%%%%%%%%%%%%%%%%%%%%%%%%%%%%%%%%%%%%%%%%%%%%%%%%%%%%%%%%%%%%%%%%%%%%%%%%%%%%%%%%%%%%%%%%%%%%%%%%%
\section{Results}\label{sec:results}
%%%%%%%%%%%%%%%%%%%%%%%%%%%%%%%%%%%%%%%%%%%%%%%%%%%%%%%%%%%%%%%%%%%%%%%%%%%%%%%%%%%%%%%%%%%%%%%%%%%%%%%%%%%%%%%%%%%%%%%%%%%%%%%%%%%%%%%%%%%%%%%%%%%%%%%%%%%%%%%%%%%%%%%%%%%%%%%%%%%%%%%%%%%%%%%%%%%%%%%%%%%%%%%%%%%%%%%%%%%%%%%%%%%

%----------------------------------------------------------------------------------------------------------------------------------%
\begin{figure}[ht!]
    \centering
    \includegraphics[width=1.0\hsize]{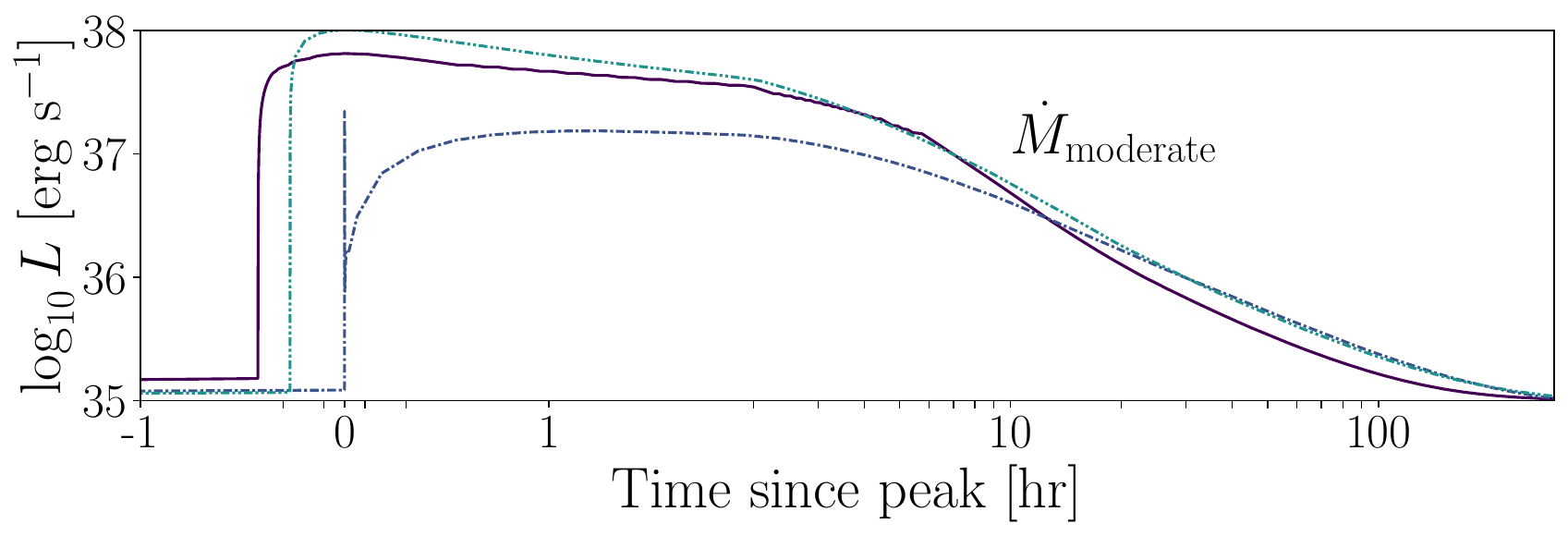}
    \includegraphics[width=1.0\hsize]{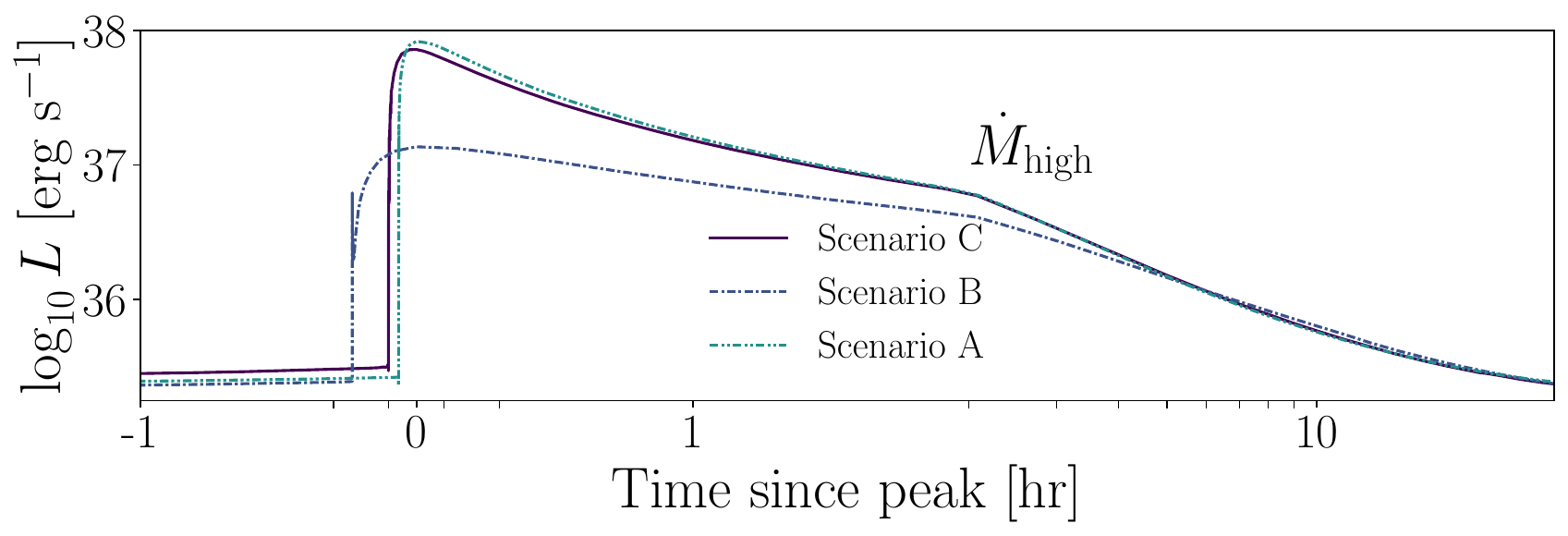}
    \caption{Superburst lightcurves for moderate (left panel) and high (right panel) accretion rates. Time has been shifted in order to set the maximum of the lightcurve at $t=0$.}
    \label{fig:lightcurves_1}
\end{figure}
%----------------------------------------------------------------------------------------------------------------------------------%

%----------------------------------------------------------------------------------------------------------------------------------%
\begin{figure}[ht!]
%    \sidecaption
    \includegraphics[width=0.9\hsize]{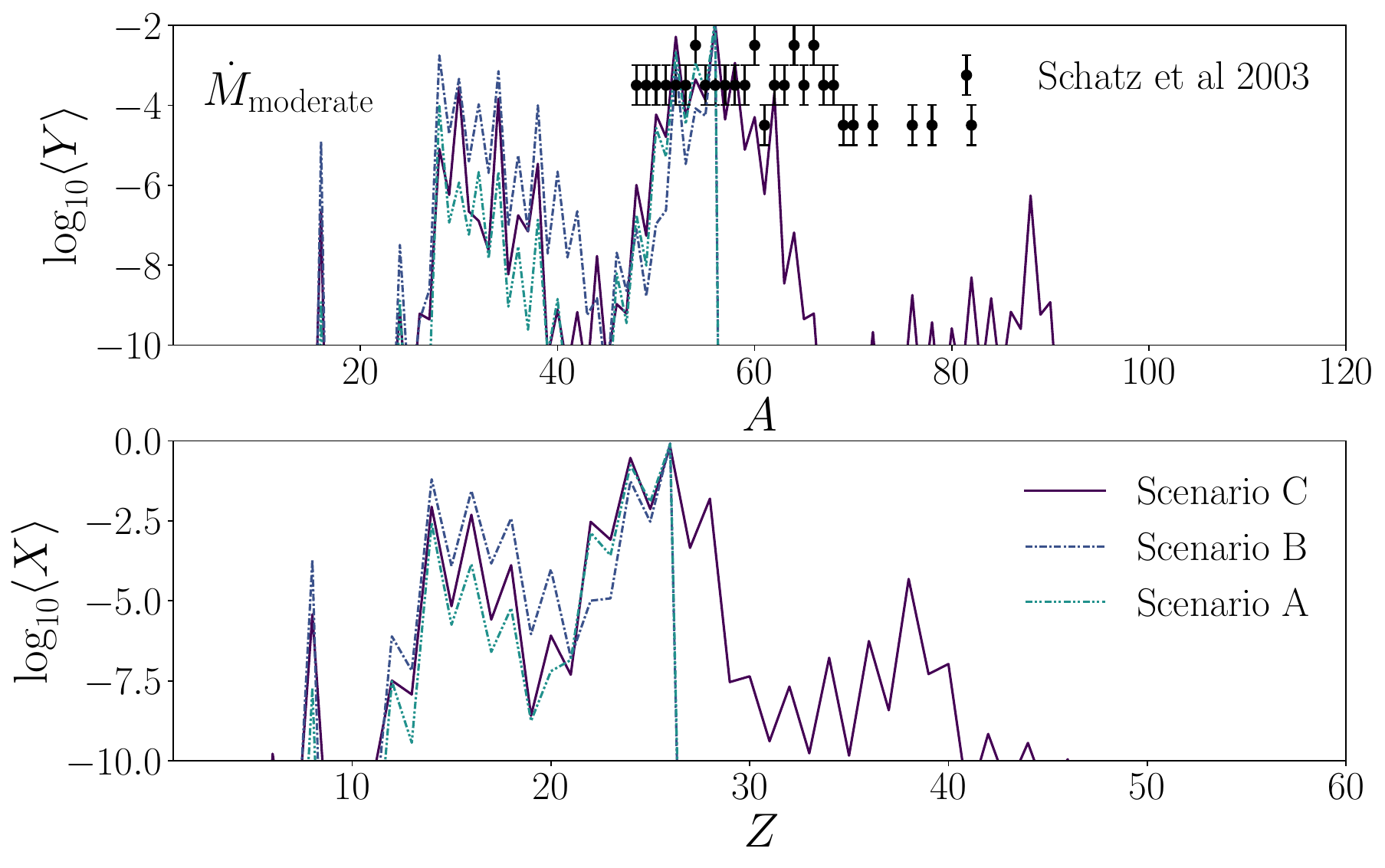}
    \caption{Distribution of ashes from the superburst for the three simulations at \middmdot. In the upper panels we include the distribution of ashes reported in \citet{Schatz_2003}.}
    \label{fig:ashes_superburst_1}
\end{figure}
%----------------------------------------------------------------------------------------------------------------------------------%

% 1. Introduction of the material.
Figure~\ref{fig:lightcurves_1} shows the superburst luminosity curves, with their main physical properties summarized in Table~\ref{tab:table_1_superbursts_parameters} in Appendix~\ref{appendix:sb_e_table}. The total released energy $E_{\rm sb}$ is obtained by integrating the luminosity from the beginning to the end of the explosion. On the one hand, the superburst properties are systematically lower in Scenario A than in Scenario C.
On the other hand, the luminosity properties of Scenarios B and C are similar. 
Similarly to KH11 and NC26, we find $y_\mathrm{ignition}\approx 10^{11}$ g cm$^{-2}$ for all our superburst simulations. These ignition depths are $\sim 1$ order of magnitude smaller than those reported for the classic fits from \citet{Cumming_2004, Cumming_2006}.

Figure~\ref{fig:ashes_superburst_1} compares the average abundances as a function of mass and charge number for the simulations at \middmdot{} between Scenarios B and C. For comparison, we also show the superburst ashes obtained with the OZM of \citet{Schatz_2003}. Our simulations display two major peaks, with local maxima at \silicos{} and \iroh{}. A third prominent peak develops around $^{88}$Sr in the Scenario C. This feature is absent in Scenario B, and is independent of the selected mass accretion rate (see Appendix ~\ref{appendix:ashes_highmdot}). Since the initial composition of the scaled rp-process simulation is already enriched in nuclei with $A=80$-$100$ (see Fig.~\ref{fig:ashes_initial}), the $^{88}$Sr peak is not due to carbon burning but rather from the processing of these pre-existing heavy nuclei. This interpretation is consistent with the pronounced depletion between the $A=56$ and $A=88$ regions, as well as with the enhanced abundance around $A=56$ compared with the relatively small abundance around $A=88$.

%%%%%%%%%%%%%%%%%%%%%%%%%%%%%%%%%%%%%%%%%%%%%%%%%%%%%%%%%%%%%%%%%%%%%%%%%%%%%%%%%%%%%%%%%%%%%%%%%%%%%%%%%%%%%%%%%%%%%%%%%%%%%%%%%%%%%%%%%%%%%%%%%%%%%%%%%%%%%%%%%%%%%%%%%%%%%%%%%%%%%%%%%%%%%%%%%%%%%%%%%%%%%%%%%%%%%%%%%%%%%%%%%%%
%%%%%%%%%%%%%%%%%%%%%%%%%%%%%%%%%%%%%%%%%%%%%%%%%%%%%%%%%%%%%%%%%%%%%%%%%%%%%%%%%%%%%%%%%%%%%%%%%%%%%%%%%%%%%%%%%%%%%%%%%%%%%%%%%%%%%%%%%%%%%%%%%%%%%%%%%%%%%%%%%%%%%%%%%%%%%%%%%%%%%%%%%%%%%%%%%%%%%%%%%%%%%%%%%%%%%%%%%%%%%%%%%%%

%%%%%%%%%%%%%%%%%%%%%%%%%%%%%%%%%%%%%%%%%%%%%%%%%%%%%%%%%%%%%%%%%%%%%%%%%%%%%%%%%%%%%%%%%%%%%%%%%%%%%%%%%%%%%%%%%%%%%%%%%%%%%%%%%%%%%%%%%%%%%%%%%%%%%%%%%%%%%%%%%%%%%%%%%%%%%%%%%%%%%%%%%%%%%%%%%%%%%%%%%%%%%%%%%%%%%%%%%%%%%%%%%%%
%%%%%%%%%%%%%%%%%%%%%%%%%%%%%%%%%%%%%%%%%%%%%%%%%%%%%%%%%%%%%%%%%%%%%%%%%%%%%%%%%%%%%%%%%%%%%%%%%%%%%%%%%%%%%%%%%%%%%%%%%%%%%%%%%%%%%%%%%%%%%%%%%%%%%%%%%%%%%%%%%%%%%%%%%%%%%%%%%%%%%%%%%%%%%%%%%%%%%%%%%%%%%%%%%%%%%%%%%%%%%%%%%%%
\section{Discussion}\label{sec:discussion}
%%%%%%%%%%%%%%%%%%%%%%%%%%%%%%%%%%%%%%%%%%%%%%%%%%%%%%%%%%%%%%%%%%%%%%%%%%%%%%%%%%%%%%%%%%%%%%%%%%%%%%%%%%%%%%%%%%%%%%%%%%%%%%%%%%%%%%%%%%%%%%%%%%%%%%%%%%%%%%%%%%%%%%%%%%%%%%%%%%%%%%%%%%%%%%%%%%%%%%%%%%%%%%%%%%%%%%%%%%%%%%%%%%%

%----------------------------------------------------------------------------------------------------------------------------------%
\begin{figure}[ht!]
    \centering
    \includegraphics[width=1.0\hsize]{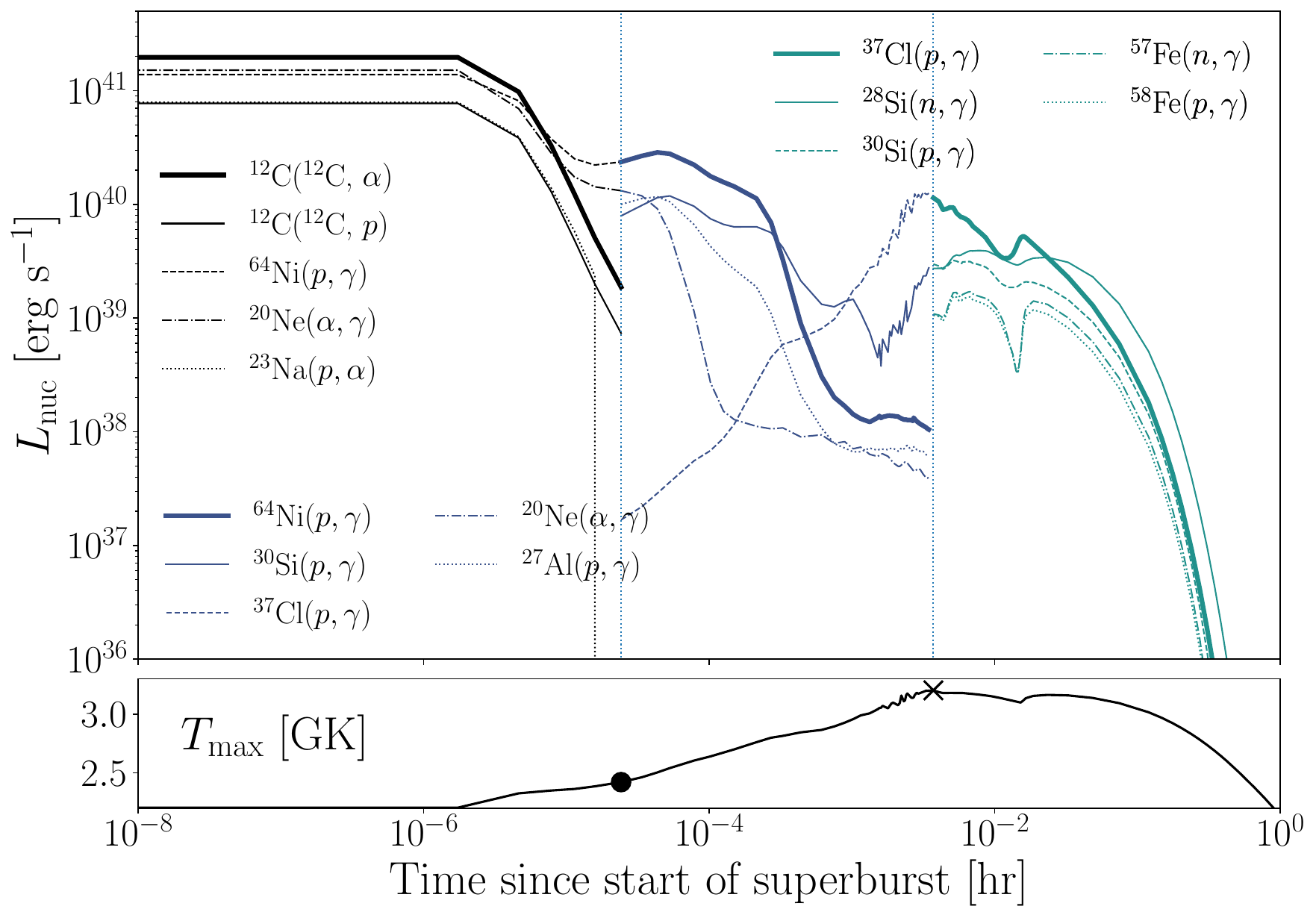}
    \caption{Net luminosity from several reactions (upper panel) and maximum temperature (lower panel) as a function of time during the superburst of Scenario C at \highmdot. Time has been shifted to start at the onset of the explosion. The total luminosity is shown by the black curve in the upper panel. The dot indicates the time at which the maximum temperature reaches $75\%$ of its absolute maximum, while the cross marks the time of the absolute maximum temperature.}
    \label{fig:nucleosynthesis_sb}
\end{figure}
%----------------------------------------------------------------------------------------------------------------------------------%

%----------------------------------------------------------------------------------------------------------------------------------%
\begin{figure}[ht!]
    \centering
    \includegraphics[width=1.0\hsize]{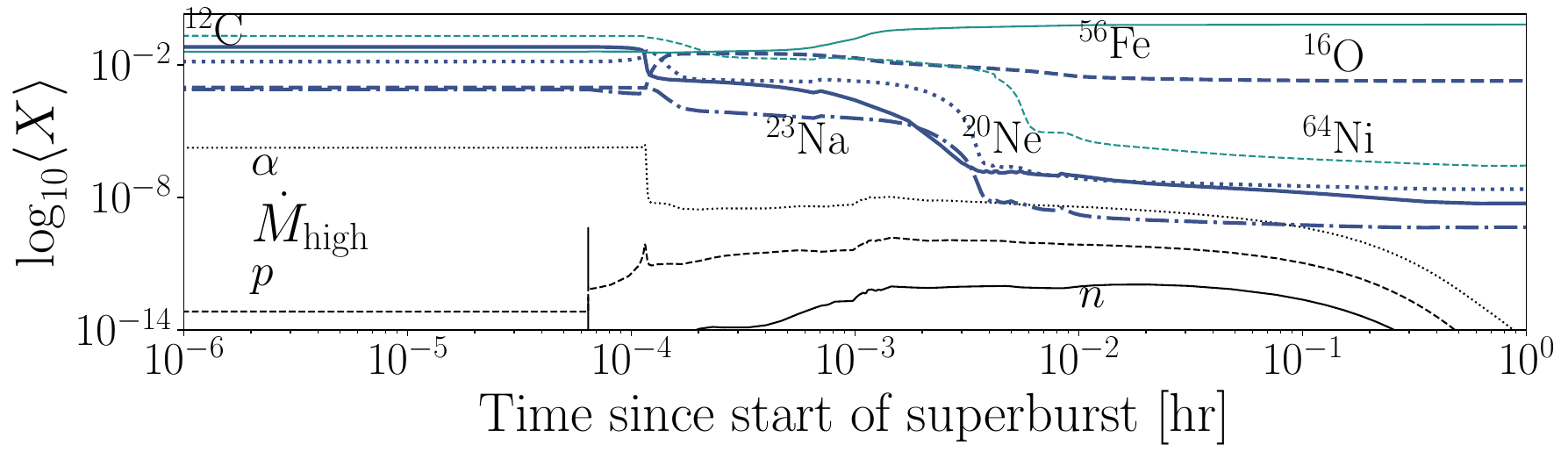}
    \caption{Average mass fraction as a function of time for the superburst of Scenario C at \highmdot. Time has been shifted to start at the onset of the explosion.}
    \label{fig:nucleosynthesis_sb_2}
\end{figure}
%----------------------------------------------------------------------------------------------------------------------------------%

%----------------------------------------------------------------------------------------------------------------------------------%
\begin{figure}[ht!]
    \centering
    \includegraphics[width=1.0\hsize]{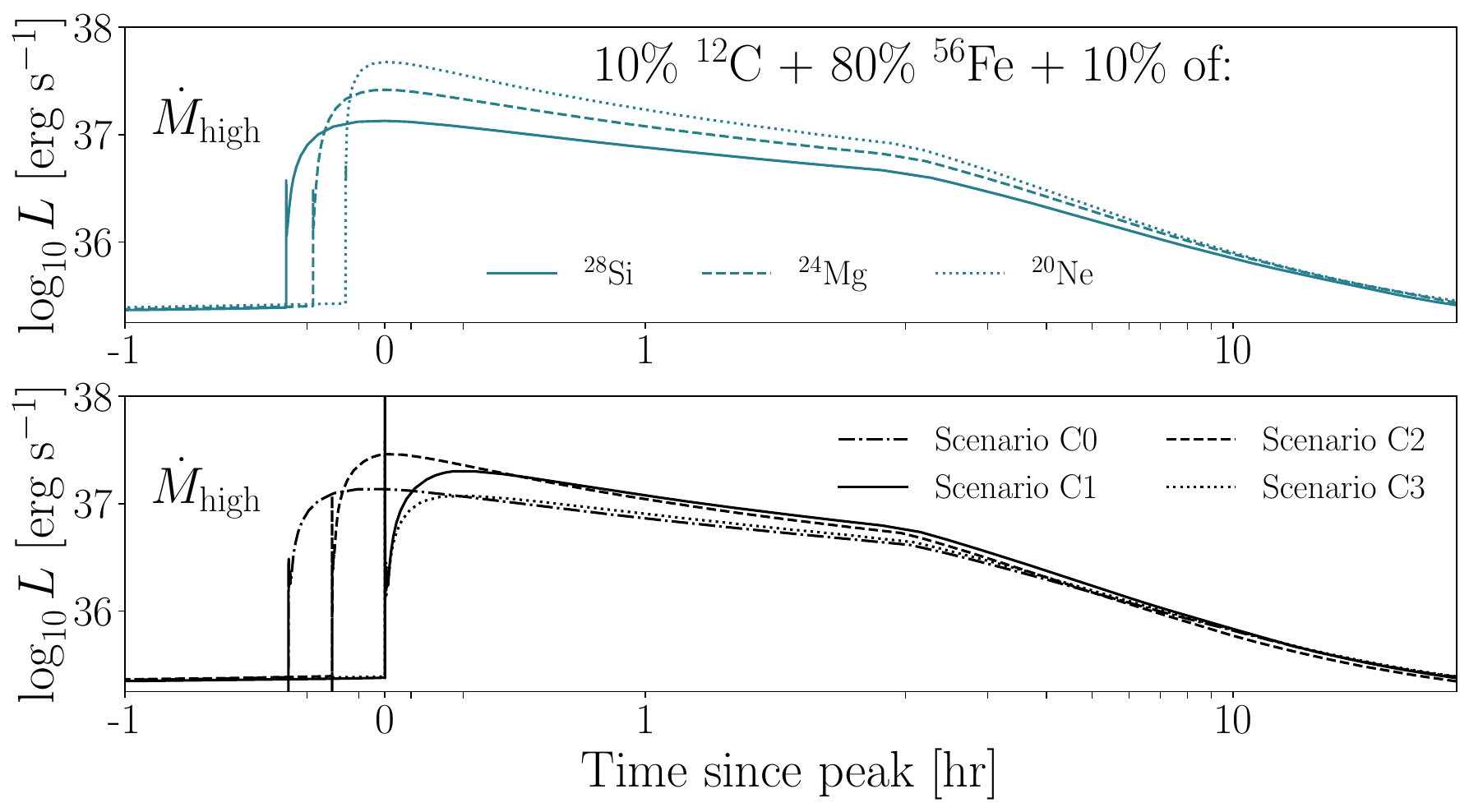}
    \caption{Upper panel: Superbursts lightcurves at \highmdot{} with different combinations of $\alpha$ nuclides and with a \carbos{} mass fraction of $10\%$.}
    \label{fig:alfanuclides_sb}
\end{figure}
%----------------------------------------------------------------------------------------------------------------------------------%

By using a scaled mass distribution from the rp-process ashes, with a \carbos{} mass fraction of only $10\%$ (Scenario C) we obtain superbursts with energy characteristics comparable to those obtained with $20\%$ \carbos{} and $80\%$ of $^{56}$Fe (Scenario B). This raises the question: which reactions, in addition to carbon fusion, provide the energy required to power the explosion? 
To address this question, in the upper panel of Fig.~\ref{fig:nucleosynthesis_sb} we plot the net luminosity associated with a few selected thermonuclear reactions\footnote{For a definition of the net luminosity, see Appendix~\ref{appendix:net_luminosity}.}, while in Fig.~\ref{fig:nucleosynthesis_sb_2} we display the evolution of the average mass fraction for a few relevant nuclides. In both figures, time is measured relative to the onset of the explosion, defined as the time at which $dL/dt = 0.1\text{max}[dL/dt]$, where $L$ is the total luminosity of the neutron star. We identify three stages after the onset of the explosion: the first stage, $t\leq 10^{-4}$ h, the runaway, in which temperature rises to $2.4$ GK, $\sim 75\%$ of the absolute maximum value; the second stage, when the temperature goes from $75\%$ to $100\%$ of the global maximum, $t\sim10^{-2}$ h after the start of the explosion. Finally, the third stage - the cooling phase.

During the first stage, carbon fusion contributes significantly to the nuclear energy generation, and a small precursor appears in the net luminosity. Although the $\alpha$-channel of carbon fusion releases more energy than the $p$-channel, the latter has important consequences for the subsequent nucleosynthesis. In particular, the rapid $(p,\alpha)$ conversion of $^{23}$Na contributes to the building of \neos{} already produced by $^{12}$C($^{12}$C,$\alpha$) and reopens a path toward the highly energetic \neos($\alpha,\gamma$)$^{24}$Mg reaction. A comparable contribution comes from  $^{64}$Ni($p,\gamma$)$^{65}$Cu, enabled by the large initial abundance of $A=64$ nuclei in the accreted material, which accounts for $\sim 30\%$ of the total mass (see Fig.~\ref{fig:ashes_initial}).
During the second stage, numerous $(p,\gamma)$ and $(n,\gamma)$ reactions are responsible for nuclear luminosities of $\sim10^{40}$ erg s$^{-1}$, albeit most of these are counterbalanced by their inverse reactions.
Some $(p,\gamma)$ reactions, such as $^{35}$Cl($p,\gamma$)$^{36}$Ar, involving nuclei near $\alpha$-nuclides between $A=28$ and $A=38$ provide a net positive luminosity. Their luminosity is one to two orders of magnitude smaller than the one coming from carbon fusion, indicating that no new dominant ignition reaction develops as the temperature rises. 
During the final stage, some $p$- or $n$-captures on heavy-$Z$ nuclides, in particular in the iron-group, provide positive net nuclear luminosities although none in excess of $10^{40}$ erg s$^{-1}$.

Given the apparent importance of other $\alpha$-nuclides for the superburst ignition, we also tested whether a composition of $10\%$ \carbos{}, $80\%$ \iroh{} and $10\%$ of another $\alpha$-nuclide can produce energetic explosions. The upper panel of Fig.~\ref{fig:alfanuclides_sb} shows the resulting lightcurves of these simulations. Consistent with the results in Fig.~\ref{fig:nucleosynthesis_sb}, the presence of $^{20}$Ne is critical for obtaining an energetic explosion. Accretion of $10\%$ of $^{24}$Mg or $^{28}$Si produces superbursts with peak luminosities $\sim0.5$ and $\sim1$ order of magnitude lower, respectively, than those obtained with $10\%$ of \neos.

The energetic contribution of $^{64}$Ni($p,\gamma$)$^{65}$Cu during the first stages of the superburst (Fig.~\ref{fig:nucleosynthesis_sb}) deserves further examination. Since the scaled rp-process ashes already contain $\sim30\%$ of $^{64}$Zn, as well as $\sim 15\%$ of $^{60}$Ni and $\sim 10\%$ of $^{68}$Ge, one might ponder what is the contribution of such abundant heavy species to the superbust energetics.
To address this question, we run four variants of Scenario C, all including $10\%$ of \carbos. In the first simulation, Scenario C0, the NS accretes $10\%$ of $^{64}$Ni and $80\%$ of \iroh. In the second one, Scenario C1, it accretes $30\%$ of $^{64}$Ni and $60\%$ of \iroh{} and in the third one, Scenario C2, it accretes 
$30\%$ of $^{64}$Ni, $15\%$ of $^{60}$Ni, $10\%$ of $^{68}$Ge and 35\% of \iroh. Finally, in Scenario C3 the NS accretes $60\%$ of $^{56}$Fe, $30\%$ of $^{64}$Zn where in addition we artificially cancel out the electron capture of $^{64}$Zn to ensure it does not decay back into $^{64}$Ni before the burst. 
Their lightcurves are also displayed in Fig.~\ref{fig:alfanuclides_sb}. 
Among the four Scenarios, C0 produces the most luminous explosion.
The Scenario C1 produces a strong precursor, reaching $L_\mathrm{max}\sim10^{39}$ erg s$^{-1}$, followed by a superburst with a comparable lightcurve as to that obtained with the simulation using the scaled rp-process ashes, Scenario C. The Scenario C2 produces a luminosity whose peak only differs down in $\sim 0.5$ (in $\log_{10}L$) with respect to the Scenario C. Finally, Scenario C3 produces a precursor but is the least luminous among the four C0, C1, C2 and C3.
The late stage of the superburst, however, is very similar among the Scenarios C1 and C2. Therefore, the enhanced abundance of heavy-$Z$ nuclides, in particular of $A = 60, 64$ and 68 appears to be as important for powering the superburst as the presence of low-$Z$ $\alpha$-nuclides. 

With these results, one might wonder why the presence of $30\%$ of $^{64}$Ni produces energetic superbursts but not the presence of $90\%$ \iroh{}? One explanation can be the difference in the $(p,\gamma)$ rate. Whereas $^{56}$Fe($p,\gamma$) releases 6.028 MeV, $^{64}$Ni($p,\gamma$) $Q$-value amounts to 7.453 MeV. Second, at the superburst temperatures above 1~GK, the rate of this latter reaction can be twice as high as that of the former. In contrast, $^{56}$Fe($p,\gamma$) reaction rate and associated heat release are found to be larger than other $(p,\gamma)$ cases in the neighbouring $A = 60, 64$ and $68$ group, such as $^{60}$Ni$(p,\gamma)$, $^{64}$Zn$(p,\gamma)$ and $^{68}$Ge$(p,\gamma)$, as shown in Appendix~\ref{appendix:rates_ratio}.
This also explains why we observe essentially the same superburst unfolding when the NS accretes $30\%$ of $^{64}$Ni plus $60\%$ of \iroh{}, Scenario C1, when compared with the $30\%$ of $^{64}$Ni plus $15\%$ of $^{60}$Ni and $10\%$ of $^{68}$Ge mixture, Scenario C2. In addition, this can also explain why we obtain a superburst with of small $L_\mathrm{max}$ when the $^{64}$Zn electron capture is forbidden, Scenario C3, in Fig.~\ref{fig:alfanuclides_sb}.

\begin{figure}[ht!]
    \centering
    \includegraphics[width=1.0\hsize]{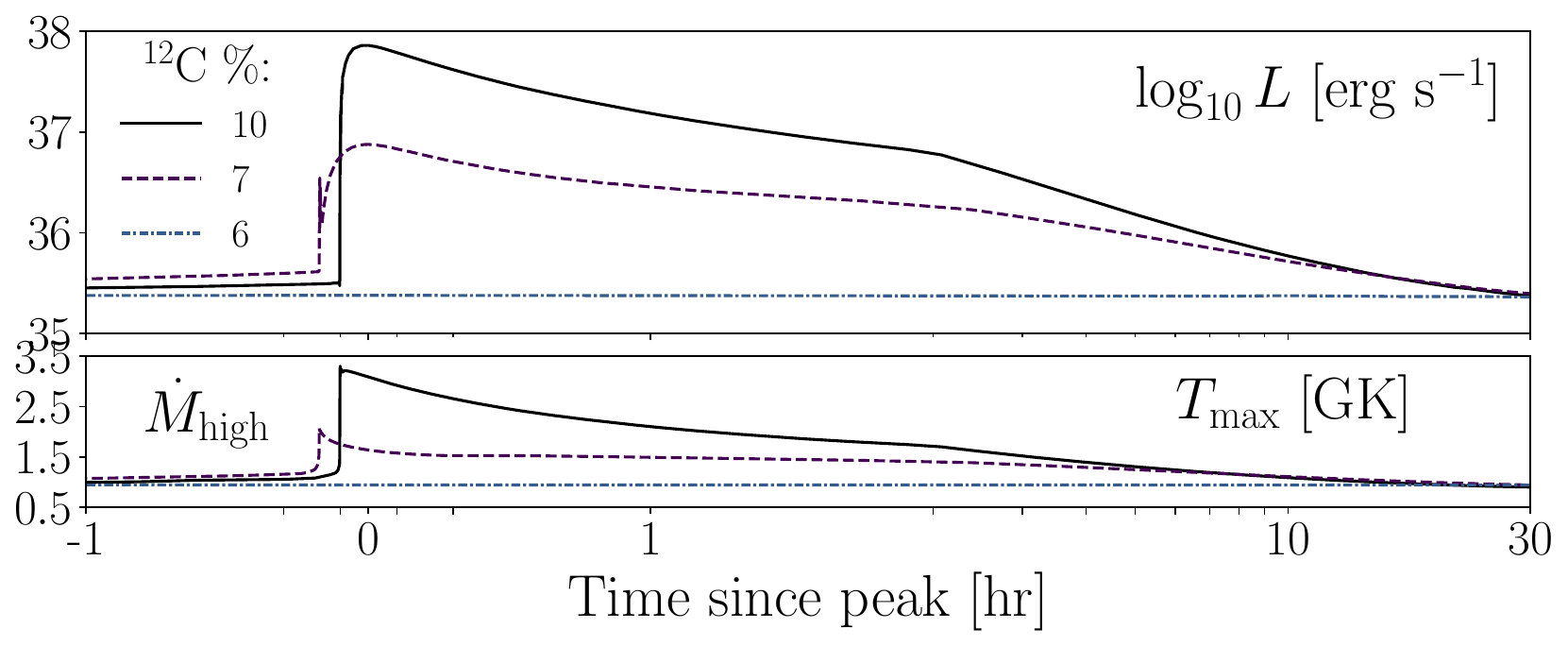}
    \caption{Top panel: Superburst lightcurves at \highmdot{} with different \carbos{} mass fraction and an accordingly scaled distribution of rp-process ashes. Bottom panel: corresponding evolution of the maximum temperature of the envelope.}
    \label{fig:xc12_lightcurves}
\end{figure}

% C12 mass fraction
Since we artificially enhanced the \carbos{} mass fraction to $10\%$ to trigger the superburst, it remains unclear what would be the minimum mass fraction of \carbos{} required to power a superbursts. To address this question, we ran additional simulations varying the \carbos{} mass fraction and accordingly re-scaling the rest of rp-process ashes. Under the same astrophysical conditions, i.e. $\dot{M}$ and $L_b$ used in our main simulations, we find that a mass fraction of $7\%$ of \carbos{} leads to stable burning at \middmdot{} whereas at \highmdot{} a superburst is obtained, with $L_\mathrm{max} = 7.6\times 10^{36}$ erg s$^{-1}$ (comparable to the least powerful explosions from Fig.~\ref{fig:alfanuclides_sb}).
Since the pre-burst temperature is comparable between the 10 and 7\% of \carbos{} simulations at \highmdot{}, as illustrated in the lower panel of Fig.~\ref{fig:xc12_lightcurves}, we can argue that the mass fraction has a more prominent role here in explaining the absence of an explosion. Recalling $\varepsilon_{^{12}\mathrm{C}}\propto Y^{2}_{^{12}\mathrm{C}}N_{A}\langle \sigma v\rangle_{T}$ - where $Y_{^{12}\mathrm{C}}$ is the molar fraction of \carbos, $N_{A}$ is the Avogadro number and $\langle \sigma v\rangle_{T}$ is the thermal \carbos{} fusion reaction rate - we see that, at the same temperature, decreasing the \carbos{}  mass fraction from $7\%$ to $6\%$ produces a change in the amount of heat released by 1.36, e.g. $\sim 40\%$.
Considering we have run only a few simulations, albeit with values for the astrophysical parameters $L_b$ and $\dot{M}$ which are consistent with existent literature of superbursts \citep[e.g.][]{KH2011, KH2012, intZand_2017, NC_2026A&A}, we set a conservative lower bound for the amount of accreted carbon required for triggering superbursts at $X(^{12}\mathrm{C}) > 7\%$.

%----------------------------------------------------------------------------------------------------------------------------------%
%----------------------------------------------------------------------------------------------------------------------------------%

%%%%%%%%%%%%%%%%%%%%%%%%%%%%%%%%%%%%%%%%%%%%%%%%%%%%%%%%%%%%%%%%%%%%%%%%%%%%%%%%%%%%%%%%%%%%%%%%%%%%%%%%%%%%%%%%%%%%%%%%%%%%%%%%%%%%%%%%%%%%%%%%%%%%%%%%%%%%%%%%%%%%%%%%%%%%%%%%%%%%%%%%%%%%%%%%%%%%%%%%%%%%%%%%%%%%%%%%%%%%%%%%%%%
%%%%%%%%%%%%%%%%%%%%%%%%%%%%%%%%%%%%%%%%%%%%%%%%%%%%%%%%%%%%%%%%%%%%%%%%%%%%%%%%%%%%%%%%%%%%%%%%%%%%%%%%%%%%%%%%%%%%%%%%%%%%%%%%%%%%%%%%%%%%%%%%%%%%%%%%%%%%%%%%%%%%%%%%%%%%%%%%%%%%%%%%%%%%%%%%%%%%%%%%%%%%%%%%%%%%%%%%%%%%%%%%%%%

%%%%%%%%%%%%%%%%%%%%%%%%%%%%%%%%%%%%%%%%%%%%%%%%%%%%%%%%%%%%%%%%%%%%%%%%%%%%%%%%%%%%%%%%%%%%%%%%%%%%%%%%%%%%%%%%%%%%%%%%%%%%%%%%%%%%%%%%%%%%%%%%%%%%%%%%%%%%%%%%%%%%%%%%%%%%%%%%%%%%%%%%%%%%%%%%%%%%%%%%%%%%%%%%%%%%%%%%%%%%%%%%%%%
%%%%%%%%%%%%%%%%%%%%%%%%%%%%%%%%%%%%%%%%%%%%%%%%%%%%%%%%%%%%%%%%%%%%%%%%%%%%%%%%%%%%%%%%%%%%%%%%%%%%%%%%%%%%%%%%%%%%%%%%%%%%%%%%%%%%%%%%%%%%%%%%%%%%%%%%%%%%%%%%%%%%%%%%%%%%%%%%%%%%%%%%%%%%%%%%%%%%%%%%%%%%%%%%%%%%%%%%%%%%%%%%%%%
\section{Conclusions}\label{sec:conclusions}
%%%%%%%%%%%%%%%%%%%%%%%%%%%%%%%%%%%%%%%%%%%%%%%%%%%%%%%%%%%%%%%%%%%%%%%%%%%%%%%%%%%%%%%%%%%%%%%%%%%%%%%%%%%%%%%%%%%%%%%%%%%%%%%%%%%%%%%%%%%%%%%%%%%%%%%%%%%%%%%%%%%%%%%%%%%%%%%%%%%%%%%%%%%%%%%%%%%%%%%%%%%%%%%%%%%%%%%%%%%%%%%%%%%

% - - - - - - - - - - - - - - - - - - - - - - - - - - - - - - - - - - - - - - - - - - - - - - - - - - - - - %
In the present work, we have taken a first step toward re-examining the superburst phenomenology by considering a more realistic fuel composition of the accreted material based on a  scaled distribution of rp-process ashes. Our results demonstrate that the composition of the ashes can play a more important role in the development of superbursts than previously expected. In particular, the presence of rp-process ashes provides additional sources of energy during the ignition and burning phases through photodisintegrations and corresponding $p$-, $\alpha$- and $n$- recaptures. As a consequence, the amount of \carbos{} required to produce energetic superbursts can be reduced from the previously considered $20\%$ to $10\%$. We further find a conservative lower bound of $\sim7\%$ for the critical \carbos{} mass fraction required to trigger a superburst.

The effect of the rp-process ashes also modifies the nucleosynthesis path and the resulting composition of the NS surface. In addition to the well-known abundance peaks around $^{28}$Si and $^{56}$Fe, we find a smaller third peak around $^{88}$Sr, which emerges from the photodisintegration of nuclei initially present in the rp-process ashes. This provides an additional signature of the influence of the initial composition on the nucleosynthesis occurring during a superburst.

In our analysis of the nuclear energetics, besides the $\alpha$- and $p$-channels of carbon fusion we find important contributions from $^{20}$Ne($\alpha,\gamma$)$^{24}$Mg and $^{64}$Ni($p,\gamma$)$^{65}$Cu. Motivated by these results, we performed additional simulations to explore whether these individual compositional components could themselves provide an alternative source of superburst energy. We find that similar energetics can be obtained either by accreting material $10\%$ enriched in $^{20}$Ne or by incorporating $\sim 55\%$ of heavy-$Z$ nuclei with $A=60$, $64$, and $68$.

Taken together, these results indicate that the composition of the accreted material — and not solely its \carbos{} content — can have a substantial impact on superburst ignition and nucleosynthesis. The sensitivity of the superburst lightcurve to both low-$Z$ $\alpha$-nuclides, such as $^{20}$Ne, and large abundances of high-$Z$ nuclei, such as $^{64}$Ni, demonstrates that different compositional components can influence the energetics through distinct reaction pathways. Consequently, considering the full distribution of rp-process ashes is important for capturing the complete nucleosynthesis path during a superburst. More generally, the role of different ratios of $\alpha$-nuclides in powering superbursts, in addition to $^{12}$C, deserves further investigation.

% - - - - - - - - - - - - - - - - - - - - - - - - - - - - - - - - - - - - - - - - - - - - - - - - - - - - - %

%%%%%%%%%%%%%%%%%%%%%%%%%%%%%%%%%%%%%%%%%%%%%%%%%%%%%%%%%%%%%%%%%%%%%%%%%%%%%%%%%%%%%%%%%%%%%%%%%%%%%%%%%%%%%%%%%%%%%%%%%%%%%%%%%%%%%%%%%%%%%%%%%%%%%%%%%%%%%%%%%%%%%%%%%%%%%%%%%%%%%%%%%%%%%%%%%%%%%%%%%%%%%%%%%%%%%%%%%%%%%%%%%%%
%%%%%%%%%%%%%%%%%%%%%%%%%%%%%%%%%%%%%%%%%%%%%%%%%%%%%%%%%%%%%%%%%%%%%%%%%%%%%%%%%%%%%%%%%%%%%%%%%%%%%%%%%%%%%%%%%%%%%%%%%%%%%%%%%%%%%%%%%%%%%%%%%%%%%%%%%%%%%%%%%%%%%%%%%%%%%%%%%%%%%%%%%%%%%%%%%%%%%%%%%%%%%%%%%%%%%%%%%%%%%%%%%%%

%%%%%%%%%%%%%%%%%%%%%%%%%%%%%%%%%%%%%%%%%%%%%%%%%%%%%%%%%%%%%%%%%%%%%%%%%%%%%%%%%%%%%%%%%%%%%%%%%%%%%%%%%%%%%%%%%%%%%%%%%%%%%%%%%%%%%%%%%%%%%%%%%%%%%%%%%%%%%%%%%%%%%%%%%%%%%%%%%%%%%%%%%%%%%%%%%%%%%%%%%%%%%%%%%%%%%%%%%%%%%%%%%%%
%%%%%%%%%%%%%%%%%%%%%%%%%%%%%%%%%%%%%%%%%%%%%%%%%%%%%%%%%%%%%%%%%%%%%%%%%%%%%%%%%%%%%%%%%%%%%%%%%%%%%%%%%%%%%%%%%%%%%%%%%%%%%%%%%%%%%%%%%%%%%%%%%%%%%%%%%%%%%%%%%%%%%%%%%%%%%%%%%%%%%%%%%%%%%%%%%%%%%%%%%%%%%%%%%%%%%%%%%%%%%%%%%%%
% ACKNOWLEDGEMENTS
%%%%%%%%%%%%%%%%%%%%%%%%%%%%%%%%%%%%%%%%%%%%%%%%%%%%%%%%%%%%%%%%%%%%%%%%%%%%%%%%%%%%%%%%%%%%%%%%%%%%%%%%%%%%%%%%%%%%%%%%%%%%%%%%%%%%%%%%%%%%%%%%%%%%%%%%%%%%%%%%%%%%%%%%%%%%%%%%%%%%%%%%%%%%%%%%%%%%%%%%%%%%%%%%%%%%%%%%%%%%%%%%%%%
\begin{acknowledgements}
M.N.-C. acknowledges support by the Fonds de la Recherche Scientifique-FNRS under Grant No IISN 4.4502.19. S.G. and N.C. are F.R.S.-FNRS research associates.
\end{acknowledgements}
%%%%%%%%%%%%%%%%%%%%%%%%%%%%%%%%%%%%%%%%%%%%%%%%%%%%%%%%%%%%%%%%%%%%%%%%%%%%%%%%%%%%%%%%%%%%%%%%%%%%%%%%%%%%%%%%%%%%%%%%%%%%%%%%%%%%%%%%%%%%%%%%%%%%%%%%%%%%%%%%%%%%%%%%%%%%%%%%%%%%%%%%%%%%%%%%%%%%%%%%%%%%%%%%%%%%%%%%%%%%%%%%%%%
%%%%%%%%%%%%%%%%%%%%%%%%%%%%%%%%%%%%%%%%%%%%%%%%%%%%%%%%%%%%%%%%%%%%%%%%%%%%%%%%%%%%%%%%%%%%%%%%%%%%%%%%%%%%%%%%%%%%%%%%%%%%%%%%%%%%%%%%%%%%%%%%%%%%%%%%%%%%%%%%%%%%%%%%%%%%%%%%%%%%%%%%%%%%%%%%%%%%%%%%%%%%%%%%%%%%%%%%%%%%%%%%%%%

%%%%%%%%%%%%%%%%%%%%%%%%%%%%%%%%%%%%%%%%%%%%%%%%%%%%%%%%%%%%%%%%%%%%%%%%%%%%%%%%%%%%%%%%%%%%%%%%%%%%%%%%%%%%%%%%%%%%%%%%%%%%%%%%%%%%%%%%%%%%%%%%%%%%%%%%%%%%%%%%%%%%%%%%%%%%%%%%%%%%%%%%%%%%%%%%%%%%%%%%%%%%%%%%%%%%%%%%%%%%%%%%%%%
%%%%%%%%%%%%%%%%%%%%%%%%%%%%%%%%%%%%%%%%%%%%%%%%%%%%%%%%%%%%%%%%%%%%%%%%%%%%%%%%%%%%%%%%%%%%%%%%%%%%%%%%%%%%%%%%%%%%%%%%%%%%%%%%%%%%%%%%%%%%%%%%%%%%%%%%%%%%%%%%%%%%%%%%%%%%%%%%%%%%%%%%%%%%%%%%%%%%%%%%%%%%%%%%%%%%%%%%%%%%%%%%%%%
% BIBLIOGRAPHY
%%%%%%%%%%%%%%%%%%%%%%%%%%%%%%%%%%%%%%%%%%%%%%%%%%%%%%%%%%%%%%%%%%%%%%%%%%%%%%%%%%%%%%%%%%%%%%%%%%%%%%%%%%%%%%%%%%%%%%%%%%%%%%%%%%%%%%%%%%%%%%%%%%%%%%%%%%%%%%%%%%%%%%%%%%%%%%%%%%%%%%%%%%%%%%%%%%%%%%%%%%%%%%%%%%%%%%%%%%%%%%%%%%%
\bibliographystyle{aa}
\bibliography{mesa_superbursts}
%%%%%%%%%%%%%%%%%%%%%%%%%%%%%%%%%%%%%%%%%%%%%%%%%%%%%%%%%%%%%%%%%%%%%%%%%%%%%%%%%%%%%%%%%%%%%%%%%%%%%%%%%%%%%%%%%%%%%%%%%%%%%%%%%%%%%%%%%%%%%%%%%%%%%%%%%%%%%%%%%%%%%%%%%%%%%%%%%%%%%%%%%%%%%%%%%%%%%%%%%%%%%%%%%%%%%%%%%%%%%%%%%%%
%%%%%%%%%%%%%%%%%%%%%%%%%%%%%%%%%%%%%%%%%%%%%%%%%%%%%%%%%%%%%%%%%%%%%%%%%%%%%%%%%%%%%%%%%%%%%%%%%%%%%%%%%%%%%%%%%%%%%%%%%%%%%%%%%%%%%%%%%%%%%%%%%%%%%%%%%%%%%%%%%%%%%%%%%%%%%%%%%%%%%%%%%%%%%%%%%%%%%%%%%%%%%%%%%%%%%%%%%%%%%%%%%%%

%%%%%%%%%%%%%%%%%%%%%%%%%%%%%%%%%%%%%%%%%%%%%%%%%%%%%%%%%%%%%%%%%%%%%%%%%%%%%%%%%%%%%%%%%%%%%%%%%%%%%%%%%%%%%%%%%%%%%%%%%%%%%%%%%%%%%%%%%%%%%%%%%%%%%%%%%%%%%%%%%%%%%%%%%%%%%%%%%%%%%%%%%%%%%%%%%%%%%%%%%%%%%%%%%%%%%%%%%%%%%%%%%%%
%%%%%%%%%%%%%%%%%%%%%%%%%%%%%%%%%%%%%%%%%%%%%%%%%%%%%%%%%%%%%%%%%%%%%%%%%%%%%%%%%%%%%%%%%%%%%%%%%%%%%%%%%%%%%%%%%%%%%%%%%%%%%%%%%%%%%%%%%%%%%%%%%%%%%%%%%%%%%%%%%%%%%%%%%%%%%%%%%%%%%%%%%%%%%%%%%%%%%%%%%%%%%%%%%%%%%%%%%%%%%%%%%%%
\begin{appendix}
%%%%%%%%%%%%%%%%%%%%%%%%%%%%%%%%%%%%%%%%%%%%%%%%%%%%%%%%%%%%%%%%%%%%%%%%%%%%%%%%%%%%%%%%%%%%%%%%%%%%%%%%%%%%%%%%%%%%%%%%%%%%%%%%%%%%%%%%%%%%%%%%%%%%%%%%%%%%%%%%%%%%%%%%%%%%%%%%%%%%%%%%%%%%%%%%%%%%%%%%%%%%%%%%%%%%%%%%%%%%%%%%%%%

%%--%%--%%--%%--%%--%%--%%--%%--%%--%%--%%--%%--%%--%%--%%--%%--%%--%%--%%--%%--%%--%%--%%--%%--%%--%%--%%--%%--%%--%%--%%--%%--%%--%%--%%--%%--%%--%%--%%--%%--%%--%%--%%--%%--%%--%%--%%--%%--%%--%%--%%--%%--%%--%%--%%--%%--%%
%%--%%--%%--%%--%%--%%--%%--%%--%%--%%--%%--%%--%%--%%--%%--%%--%%--%%--%%--%%--%%--%%--%%--%%--%%--%%--%%--%%--%%--%%--%%--%%--%%--%%--%%--%%--%%--%%--%%--%%--%%--%%--%%--%%--%%--%%--%%--%%--%%--%%--%%--%%--%%--%%--%%--%%--%%
\section{Calculation of the rp-process ashes}\label{appendix:rp_ashes_calcs}
%%--%%--%%--%%--%%--%%--%%--%%--%%--%%--%%--%%--%%--%%--%%--%%--%%--%%--%%--%%--%%--%%--%%--%%--%%--%%--%%--%%--%%--%%--%%--%%--%%--%%--%%--%%--%%--%%--%%--%%--%%--%%--%%--%%--%%--%%--%%--%%--%%--%%--%%--%%--%%--%%--%%--%%--%%

%----------------------------------------------------------------------------------------------------------------------------------%
\begin{figure}[ht!]
    \centering
    \includegraphics[width=1.0\hsize]{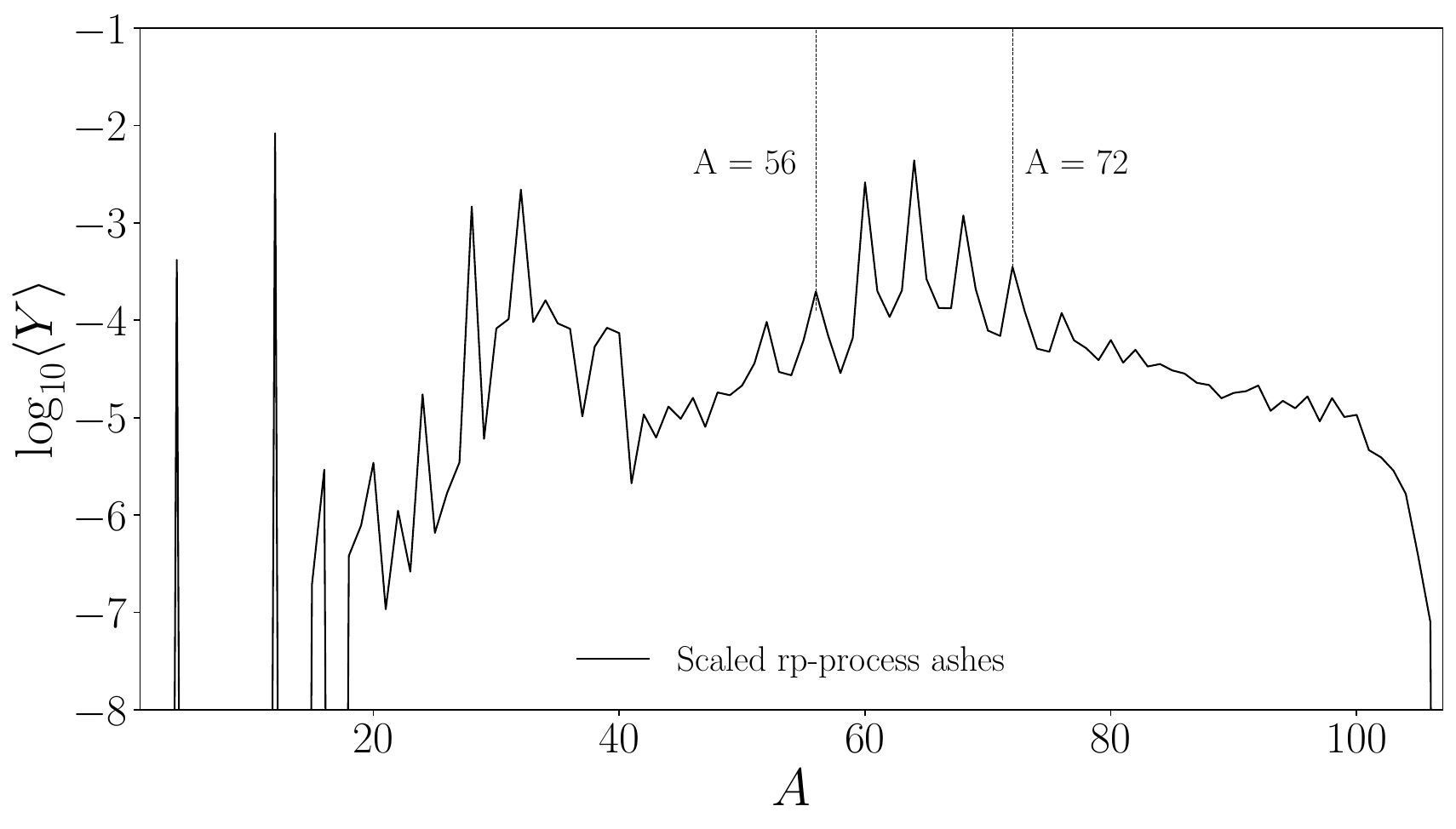}
    \caption{Scaled distribution of ashes from the rp-process obtained by selecting the 200 most abundant nuclides and artificially enhancing the $^{12}$C mass fraction to 10\%.}
    \label{fig:ashes_initial}
\end{figure}
%----------------------------------------------------------------------------------------------------------------------------------%

The distribution of ashes from the rp-process we considered in the present work was obtained from a separate calculation we performed. We ran simulations with \texttt{MESA}, accreting solar-like material ($70\%$ of \hydros) at $3\times 10^{-9}$ $M_{\odot}\, \mathrm{yr}^{-1}$ for a series of regular bursts. From their ashes, we extracted the chemical composition from a suitable average (similar to what we reported in Appendix A from NC26. Next, we renormalized this chemical composition to keep only the 200 most abundant nuclides whose mass fraction were $\geq 10^{-6}$. We then performed a second renormalization to artificially increase the mass fraction of \carbos{} to a desired value (namely $10\%$) without changing the relative proportions of the other nuclides. The resulting distribution of ashes is plotted in Figure~\ref{fig:ashes_initial}.

%%--%%--%%--%%--%%--%%--%%--%%--%%--%%--%%--%%--%%--%%--%%--%%--%%--%%--%%--%%--%%--%%--%%--%%--%%--%%--%%--%%--%%--%%--%%--%%--%%--%%--%%--%%--%%--%%--%%--%%--%%--%%--%%--%%--%%--%%--%%--%%--%%--%%--%%--%%--%%--%%--%%--%%--%%
%%--%%--%%--%%--%%--%%--%%--%%--%%--%%--%%--%%--%%--%%--%%--%%--%%--%%--%%--%%--%%--%%--%%--%%--%%--%%--%%--%%--%%--%%--%%--%%--%%--%%--%%--%%--%%--%%--%%--%%--%%--%%--%%--%%--%%--%%--%%--%%--%%--%%--%%--%%--%%--%%--%%--%%--%%

%%--%%--%%--%%--%%--%%--%%--%%--%%--%%--%%--%%--%%--%%--%%--%%--%%--%%--%%--%%--%%--%%--%%--%%--%%--%%--%%--%%--%%--%%--%%--%%--%%--%%--%%--%%--%%--%%--%%--%%--%%--%%--%%--%%--%%--%%--%%--%%--%%--%%--%%--%%--%%--%%--%%--%%--%%
%%--%%--%%--%%--%%--%%--%%--%%--%%--%%--%%--%%--%%--%%--%%--%%--%%--%%--%%--%%--%%--%%--%%--%%--%%--%%--%%--%%--%%--%%--%%--%%--%%--%%--%%--%%--%%--%%--%%--%%--%%--%%--%%--%%--%%--%%--%%--%%--%%--%%--%%--%%--%%--%%--%%--%%--%%
\section{Dependency on the size of the network}\label{appendix:network_size}
%%--%%--%%--%%--%%--%%--%%--%%--%%--%%--%%--%%--%%--%%--%%--%%--%%--%%--%%--%%--%%--%%--%%--%%--%%--%%--%%--%%--%%--%%--%%--%%--%%--%%--%%--%%--%%--%%--%%--%%--%%--%%--%%--%%--%%--%%--%%--%%--%%--%%--%%--%%--%%--%%--%%--%%--%%

%----------------------------------------------------------------------------------------------------------------------------------%
\begin{figure}[ht!]
    \centering
    \includegraphics[width=1\hsize]{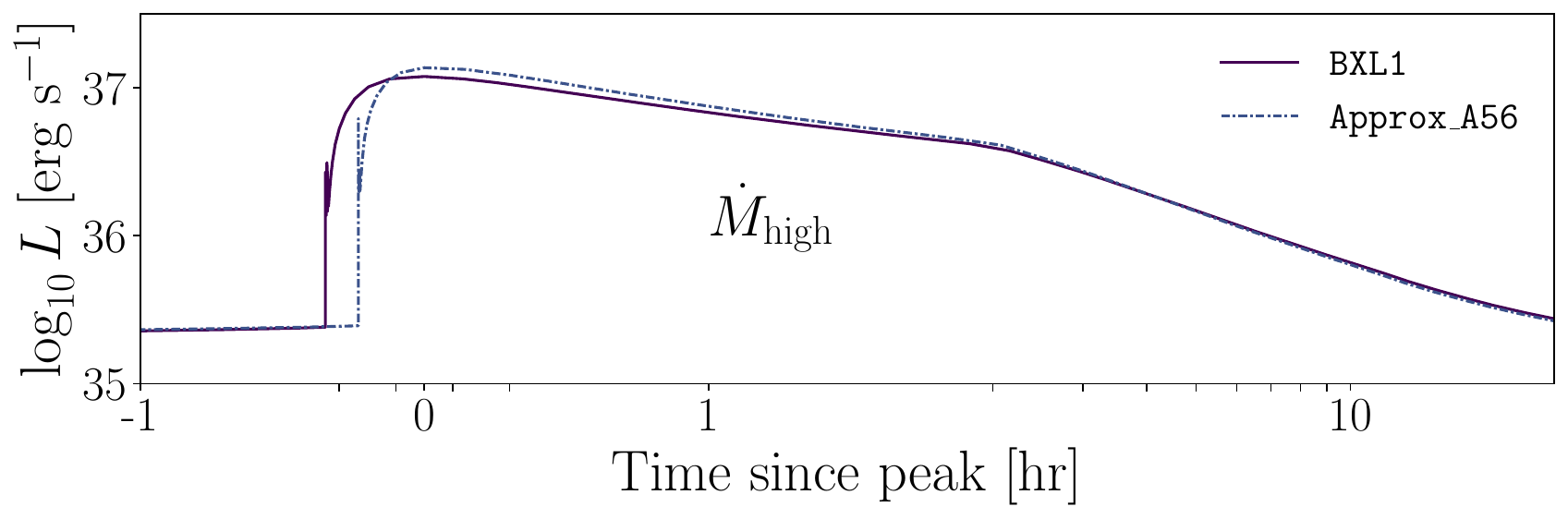}
    \caption{Superburst luminosity curves with the networks \texttt{BXL1} and \approxai.}
    \label{fig:lightcurve_bxl1_vs_bxl4}
\end{figure}
%----------------------------------------------------------------------------------------------------------------------------------%

%----------------------------------------------------------------------------------------------------------------------------------%
\begin{figure}[ht!]
    \centering
    \includegraphics[width=1\hsize]{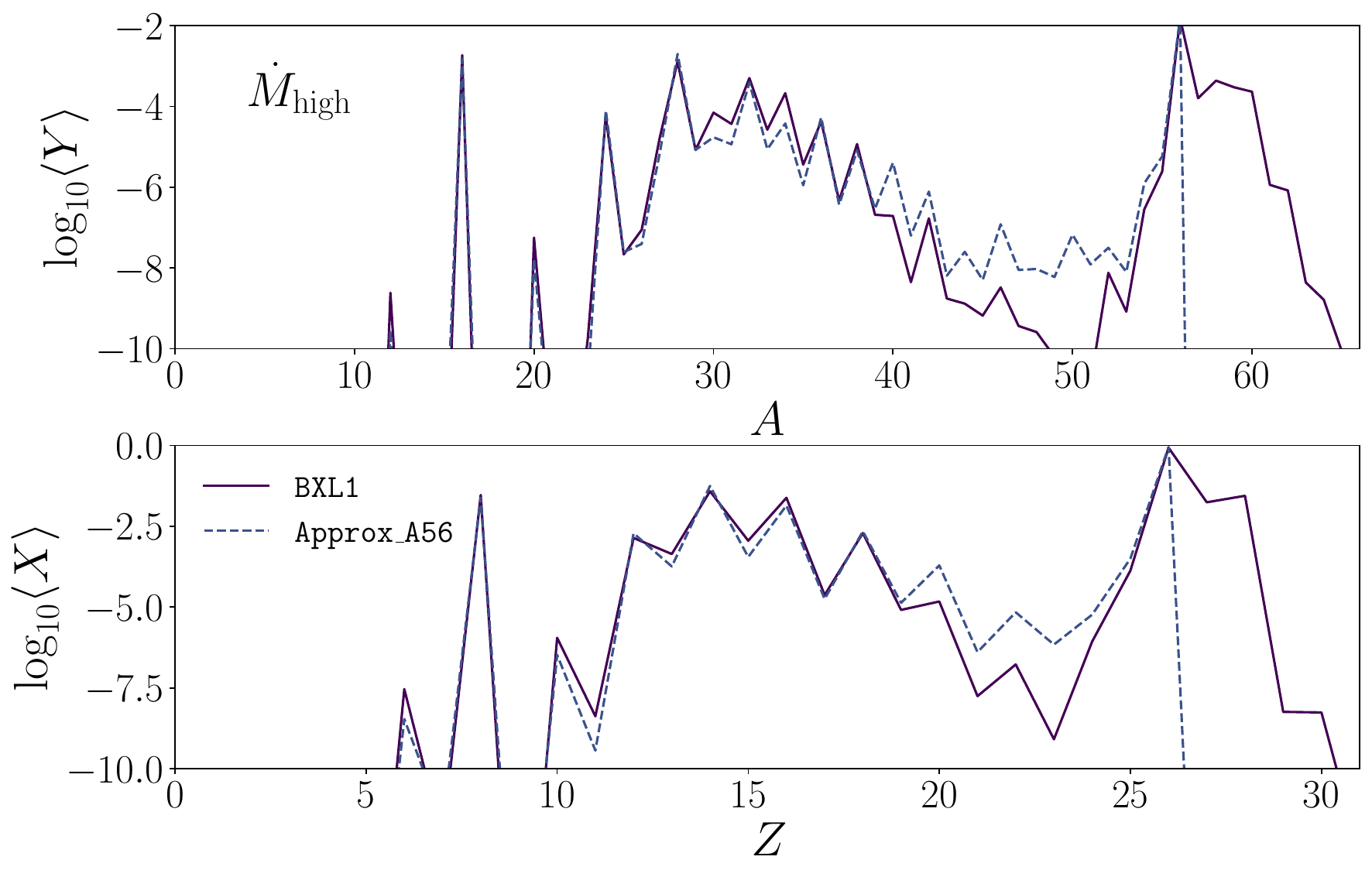}
    \caption{Distribution of ashes from the superburst with the networks \texttt{BXL1} and \approxai. In both cases, the mass fraction of \iroh{} is $90\%$.}
    \label{fig:ashes_bxl1_vs_bxl4}
\end{figure}
%----------------------------------------------------------------------------------------------------------------------------------%

In the present work, when considering a \carbos{} mass fraction of 10\% and different mass distributions for the remaining 90\%, Scenarios B and C in the main text, we compared superburst lightcurves with two different networks: one with nuclides of $A\leq 56$ and the other with nuclides of $A\leq 120$. The question therefore arises as to whether the size of the networks could affect our conclusions.

In NC26, when accreting 20\% of \carbos{} and 80\% of \iroh, we found only small differences in the superburst ashes upon increasing the size of the network from \approxai{} to \approxaii. This latter consists of nuclides in and above the valley of stability with $A\leq 64$. In particular, we observed that the nucleosynthesis barely proceeded beyond $A = 62$. Since we also observed a small difference in the lightcurves, we concluded that it is safe to use a network of reactions with $A\leq 56$ for simulating superbursts.

Nevertheless, to check that this is still the case here, we performed an additional simulation, accreting 10\% of \carbos{} and 90\% of \iroh{} using the network \bxlone{} instead of \approxai. As can be seen in Fig.~\ref{fig:lightcurve_bxl1_vs_bxl4}, we observe only a minor difference, in the order of $\sim 5\%$.

Regarding the composition of the ashes displayed in Fig.~\ref{fig:ashes_bxl1_vs_bxl4}, we observe a similar difference as to the one we identified in \citet{NC_2026A&A}. Namely, that nucleosynthesis can proceed only up to $A\sim 62$.

%%--%%--%%--%%--%%--%%--%%--%%--%%--%%--%%--%%--%%--%%--%%--%%--%%--%%--%%--%%--%%--%%--%%--%%--%%--%%--%%--%%--%%--%%--%%--%%--%%--%%--%%--%%--%%--%%--%%--%%--%%--%%--%%--%%--%%--%%--%%--%%--%%--%%--%%--%%--%%--%%--%%--%%--%%
%%--%%--%%--%%--%%--%%--%%--%%--%%--%%--%%--%%--%%--%%--%%--%%--%%--%%--%%--%%--%%--%%--%%--%%--%%--%%--%%--%%--%%--%%--%%--%%--%%--%%--%%--%%--%%--%%--%%--%%--%%--%%--%%--%%--%%--%%--%%--%%--%%--%%--%%--%%--%%--%%--%%--%%--%%

%%--%%--%%--%%--%%--%%--%%--%%--%%--%%--%%--%%--%%--%%--%%--%%--%%--%%--%%--%%--%%--%%--%%--%%--%%--%%--%%--%%--%%--%%--%%--%%--%%--%%--%%--%%--%%--%%--%%--%%--%%--%%--%%--%%--%%--%%--%%--%%--%%--%%--%%--%%--%%--%%--%%--%%--%%
%%--%%--%%--%%--%%--%%--%%--%%--%%--%%--%%--%%--%%--%%--%%--%%--%%--%%--%%--%%--%%--%%--%%--%%--%%--%%--%%--%%--%%--%%--%%--%%--%%--%%--%%--%%--%%--%%--%%--%%--%%--%%--%%--%%--%%--%%--%%--%%--%%--%%--%%--%%--%%--%%--%%--%%--%%
\section{Superburst energetic properties}\label{appendix:sb_e_table}
%%--%%--%%--%%--%%--%%--%%--%%--%%--%%--%%--%%--%%--%%--%%--%%--%%--%%--%%--%%--%%--%%--%%--%%--%%--%%--%%--%%--%%--%%--%%--%%--%%--%%--%%--%%--%%--%%--%%--%%--%%--%%--%%--%%--%%--%%--%%--%%--%%--%%--%%--%%--%%--%%--%%--%%--%%

%----------------------------------------------------------------------------------------------------------------------------------------------------------------------------------------------------------------------------%
\begin{table}[ht!]
\scriptsize
\caption{Properties of superbursts for different compositions of the accreted material.}
\label{tab:table_1_superbursts_parameters}
\centering
\setlength{\tabcolsep}{5pt}
\begin{tabular}{lccc}
\hline
\multicolumn{4}{c}{\middmdot} \\
\hline
Ashes & $L_\mathrm{max}$ & $y_\mathrm{ignition}$ & $E_\mathrm{sb}$ \\
& [$10^{37}$ erg s$^{-1}$] & [$10^{11}$ g cm$^{-2}$] & [$10^{42}$ erg] \\
\hline
(Scenario C) 10\% C + scaled rp-process ashes & 6.51 & 3.16 & 2.04 \\
(Scenario A) 20\% C + 80\% \iroh & 10.01 & 4.20 & 1.87 \\
(Scenario B) 10\% C + 90\% \iroh & 2.22 & 3.98 & 1.21 \\
\hline
\multicolumn{4}{c}{\highmdot} \\
\hline
Ashes & $L_\mathrm{max}$ & $y_\mathrm{ignition}$ & $E_\mathrm{sb}$ \\
& [$10^{37}$ erg s$^{-1}$] & [$10^{11}$ g cm$^{-2}$] & [$10^{42}$ erg] \\
\hline
(Scenario C) 10\% C + scaled rp-process ashes & 7.22 & 0.8 & 0.43 \\
(Scenario A) 20\% C + 80\% \iroh & 8.32 & 0.96 & 0.41 \\
(Scenario B) 10\% C + 90\% \iroh & 1.37 & 1.00 & 0.23 \\
\hline
\end{tabular}
\end{table}
%----------------------------------------------------------------------------------------------------------------------------------------------------------------------------------------------------------------------------%
In Table~\ref{tab:table_1_superbursts_parameters} we summarize the properties of the superbursts from our three main scenarios.
%%--%%--%%--%%--%%--%%--%%--%%--%%--%%--%%--%%--%%--%%--%%--%%--%%--%%--%%--%%--%%--%%--%%--%%--%%--%%--%%--%%--%%--%%--%%--%%--%%--%%--%%--%%--%%--%%--%%--%%--%%--%%--%%--%%--%%--%%--%%--%%--%%--%%--%%--%%--%%--%%--%%--%%--%%
%%--%%--%%--%%--%%--%%--%%--%%--%%--%%--%%--%%--%%--%%--%%--%%--%%--%%--%%--%%--%%--%%--%%--%%--%%--%%--%%--%%--%%--%%--%%--%%--%%--%%--%%--%%--%%--%%--%%--%%--%%--%%--%%--%%--%%--%%--%%--%%--%%--%%--%%--%%--%%--%%--%%--%%--%%

%%--%%--%%--%%--%%--%%--%%--%%--%%--%%--%%--%%--%%--%%--%%--%%--%%--%%--%%--%%--%%--%%--%%--%%--%%--%%--%%--%%--%%--%%--%%--%%--%%--%%--%%--%%--%%--%%--%%--%%--%%--%%--%%--%%--%%--%%--%%--%%--%%--%%--%%--%%--%%--%%--%%--%%--%%
%%--%%--%%--%%--%%--%%--%%--%%--%%--%%--%%--%%--%%--%%--%%--%%--%%--%%--%%--%%--%%--%%--%%--%%--%%--%%--%%--%%--%%--%%--%%--%%--%%--%%--%%--%%--%%--%%--%%--%%--%%--%%--%%--%%--%%--%%--%%--%%--%%--%%--%%--%%--%%--%%--%%--%%--%%
\section{Distribution of superburst ashes at \highmdot}\label{appendix:ashes_highmdot}
%%--%%--%%--%%--%%--%%--%%--%%--%%--%%--%%--%%--%%--%%--%%--%%--%%--%%--%%--%%--%%--%%--%%--%%--%%--%%--%%--%%--%%--%%--%%--%%--%%--%%--%%--%%--%%--%%--%%--%%--%%--%%--%%--%%--%%--%%--%%--%%--%%--%%--%%--%%--%%--%%--%%--%%--%%

%----------------------------------------------------------------------------------------------------------------------------------%
\begin{figure}[ht!]
    \sidecaption
    \includegraphics[width=1.0\hsize]{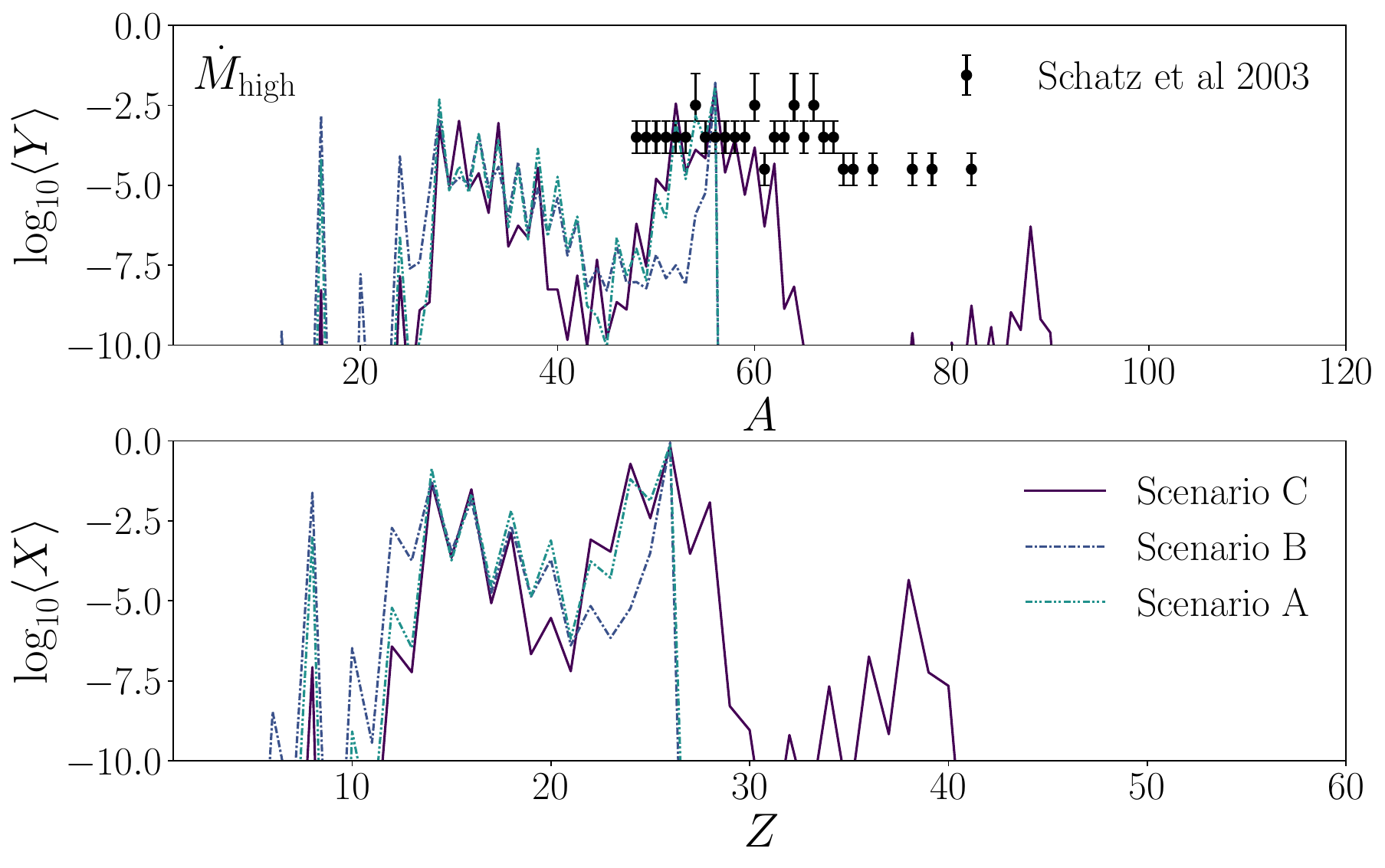}
    \caption{Distribution of ashes from the superburst for the two simulations at \highmdot{}. In the upper panels we include the distribution of ashes reported in \citet{Schatz_2003}.}
    \label{fig:ashes_superburst_2}
\end{figure}
%----------------------------------------------------------------------------------------------------------------------------------%

Since the distribution of superburst ashes is fairly similar between the simulations at \middmdot{} and at \highmdot{}, we opted for including only one of them in the main text while leaving the second one for this appendix. The corresponding distribution is presented in Fig.~\ref{fig:ashes_superburst_2}.

%%--%%--%%--%%--%%--%%--%%--%%--%%--%%--%%--%%--%%--%%--%%--%%--%%--%%--%%--%%--%%--%%--%%--%%--%%--%%--%%--%%--%%--%%--%%--%%--%%--%%--%%--%%--%%--%%--%%--%%--%%--%%--%%--%%--%%--%%--%%--%%--%%--%%--%%--%%--%%--%%--%%--%%--%%
%%--%%--%%--%%--%%--%%--%%--%%--%%--%%--%%--%%--%%--%%--%%--%%--%%--%%--%%--%%--%%--%%--%%--%%--%%--%%--%%--%%--%%--%%--%%--%%--%%--%%--%%--%%--%%--%%--%%--%%--%%--%%--%%--%%--%%--%%--%%--%%--%%--%%--%%--%%--%%--%%--%%--%%--%%

%%--%%--%%--%%--%%--%%--%%--%%--%%--%%--%%--%%--%%--%%--%%--%%--%%--%%--%%--%%--%%--%%--%%--%%--%%--%%--%%--%%--%%--%%--%%--%%--%%--%%--%%--%%--%%--%%--%%--%%--%%--%%--%%--%%--%%--%%--%%--%%--%%--%%--%%--%%--%%--%%--%%--%%--%%
%%--%%--%%--%%--%%--%%--%%--%%--%%--%%--%%--%%--%%--%%--%%--%%--%%--%%--%%--%%--%%--%%--%%--%%--%%--%%--%%--%%--%%--%%--%%--%%--%%--%%--%%--%%--%%--%%--%%--%%--%%--%%--%%--%%--%%--%%--%%--%%--%%--%%--%%--%%--%%--%%--%%--%%--%%
\section{Net luminosity}\label{appendix:net_luminosity}
%%--%%--%%--%%--%%--%%--%%--%%--%%--%%--%%--%%--%%--%%--%%--%%--%%--%%--%%--%%--%%--%%--%%--%%--%%--%%--%%--%%--%%--%%--%%--%%--%%--%%--%%--%%--%%--%%--%%--%%--%%--%%--%%--%%--%%--%%--%%--%%--%%--%%--%%--%%--%%--%%--%%--%%--%%
Let $\varepsilon_{d}>0$ and $\varepsilon_{i}>0$ be the specific energy generation rates associated with the forward nuclear reaction $d$ and its inverse reaction $i$, respectively. The corresponding luminosities are defined as
\begin{align}
L_{d} &:= \int dm\, \varepsilon_{d},\\
L_{i} &:= \int dm\, \varepsilon_{i}.
\end{align}
where the integration is performed over the whole neutron star envelope. Numerically, this corresponds to the region between $P = 10^{20}$ erg cm$^{-3}$, the surface of the star, and the envelope-crust boundary, $P \sim 10^{25}$ erg cm$^{-3}$. By definition of the specific energy generation rates, we see $L_d >0$ and $L_i>0$. Throughout the main text, the net luminosity associated with the forward reaction $d$ is defined as
\begin{equation}
L_\mathrm{nuc, net} := L_{d} - L_{i}.
\end{equation}
Regarding the physical interpretation: from the definition of the individual luminosities, whenever $L_\mathrm{nuc, net}>0$ occurs it means the forward nuclear reaction dominates over the inverse reaction. Similarly, if $L_\mathrm{nuc, net}<0$ then it means the inverse reaction dominates over the forward one.

%%--%%--%%--%%--%%--%%--%%--%%--%%--%%--%%--%%--%%--%%--%%--%%--%%--%%--%%--%%--%%--%%--%%--%%--%%--%%--%%--%%--%%--%%--%%--%%--%%--%%--%%--%%--%%--%%--%%--%%--%%--%%--%%--%%--%%--%%--%%--%%--%%--%%--%%--%%--%%--%%--%%--%%--%%
%%--%%--%%--%%--%%--%%--%%--%%--%%--%%--%%--%%--%%--%%--%%--%%--%%--%%--%%--%%--%%--%%--%%--%%--%%--%%--%%--%%--%%--%%--%%--%%--%%--%%--%%--%%--%%--%%--%%--%%--%%--%%--%%--%%--%%--%%--%%--%%--%%--%%--%%--%%--%%--%%--%%--%%--%%

%%--%%--%%--%%--%%--%%--%%--%%--%%--%%--%%--%%--%%--%%--%%--%%--%%--%%--%%--%%--%%--%%--%%--%%--%%--%%--%%--%%--%%--%%--%%--%%--%%--%%--%%--%%--%%--%%--%%--%%--%%--%%--%%--%%--%%--%%--%%--%%--%%--%%--%%--%%--%%--%%--%%--%%--%%
%%--%%--%%--%%--%%--%%--%%--%%--%%--%%--%%--%%--%%--%%--%%--%%--%%--%%--%%--%%--%%--%%--%%--%%--%%--%%--%%--%%--%%--%%--%%--%%--%%--%%--%%--%%--%%--%%--%%--%%--%%--%%--%%--%%--%%--%%--%%--%%--%%--%%--%%--%%--%%--%%--%%--%%--%%
\section{Comparison of $(p,\gamma)$ reaction rates over $^{56}$Fe, $^{60}$Ni, $^{64}$Ni, $^{64}$Zn and $^{68}$Ge.}\label{appendix:rates_ratio}
%%--%%--%%--%%--%%--%%--%%--%%--%%--%%--%%--%%--%%--%%--%%--%%--%%--%%--%%--%%--%%--%%--%%--%%--%%--%%--%%--%%--%%--%%--%%--%%--%%--%%--%%--%%--%%--%%--%%--%%--%%--%%--%%--%%--%%--%%--%%--%%--%%--%%--%%--%%--%%--%%--%%--%%--%%
%----------------------------------------------------------------------------------------------------------------------------------%
\begin{figure}[ht!]
    \centering
    \includegraphics[width=1.0\hsize]{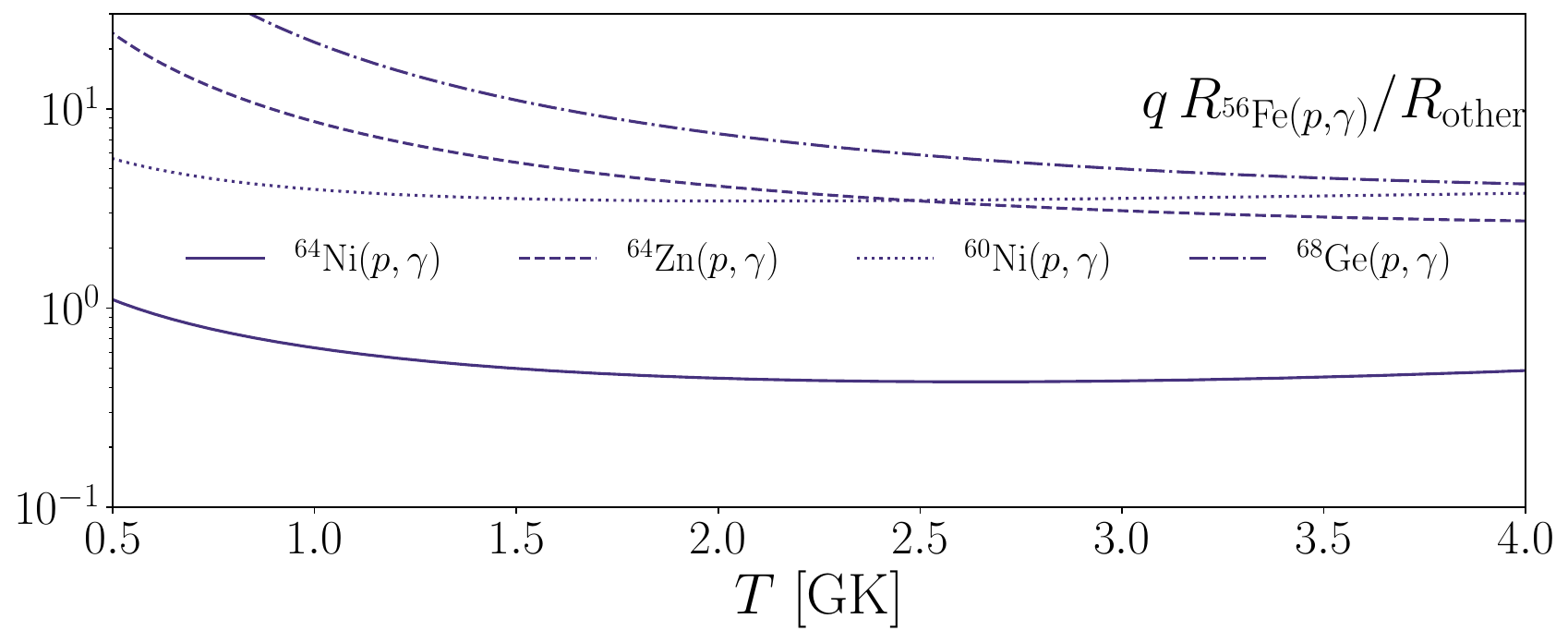}    
    \caption{Ratio of reaction rates as a function of temperature, multiplied by the ratio $q$ of their corresponding $Q$ values.}
    \label{fig:rates_ratio_1}
\end{figure}
%----------------------------------------------------------------------------------------------------------------------------------%

Fig.~\ref{fig:rates_ratio_1} illustrates the ratio of the \iroh$(p,\gamma)$ reaction rate with respect to four proton capture rates, as a function of temperature. These reaction rates are associated with several $A=60,64,68$ nuclides which are present in the ashes from the rp-process. The reactions are: $^{64}$Ni($p,\gamma$), $^{64}$Zn($p,\gamma$), $^{60}$Ni($p,\gamma$) and $^{68}$Ge($p,\gamma$). For completeness, we have multiplied the rates ratio by the ratio of their corresponding $Q$-values, which we denote by $q$.

%%--%%--%%--%%--%%--%%--%%--%%--%%--%%--%%--%%--%%--%%--%%--%%--%%--%%--%%--%%--%%--%%--%%--%%--%%--%%--%%--%%--%%--%%--%%--%%--%%--%%--%%--%%--%%--%%--%%--%%--%%--%%--%%--%%--%%--%%--%%--%%--%%--%%--%%--%%--%%--%%--%%--%%--%%
%%--%%--%%--%%--%%--%%--%%--%%--%%--%%--%%--%%--%%--%%--%%--%%--%%--%%--%%--%%--%%--%%--%%--%%--%%--%%--%%--%%--%%--%%--%%--%%--%%--%%--%%--%%--%%--%%--%%--%%--%%--%%--%%--%%--%%--%%--%%--%%--%%--%%--%%--%%--%%--%%--%%--%%--%%

\end{appendix}

%%%%%%%%%%%%%%%%%%%%%%%%%%%%%%%%%%%%%%%%%%%%%%%%%%%%%%%%%%%%%%%%%%%%%%%%%%%%%%%%%%%%%%%%%%%%%%%%%%%%%%%%%%%%%%%%%%%%%%%%%%%%%%%%%%%%%%%%%%%%%%%%%%%%%%%%%%%%%%%%%%%%%%%%%%%%%%%%%%%%%%%%%%%%%%%%%%%%%%%%%%%%%%%%%%%%%%%%%%%%%%%%%%%
%%%%%%%%%%%%%%%%%%%%%%%%%%%%%%%%%%%%%%%%%%%%%%%%%%%%%%%%%%%%%%%%%%%%%%%%%%%%%%%%%%%%%%%%%%%%%%%%%%%%%%%%%%%%%%%%%%%%%%%%%%%%%%%%%%%%%%%%%%%%%%%%%%%%%%%%%%%%%%%%%%%%%%%%%%%%%%%%%%%%%%%%%%%%%%%%%%%%%%%%%%%%%%%%%%%%%%%%%%%%%%%%%%%

%%%%%%%%%%%%%%%%%%%%%%%%%%%%%%%%%%%%%%%%%%%%%%%%%%%%%%%%%%%%%%%%%%%%%%%%%%%%%%%%%%%%%%%%%%%%%%%%%%%%%%%%%%%%%%%%%%%%%%%%%%%%%%%%%%%%%%%%%%%%%%%%%%%%%%%%%%%%%%%%%%%%%%%%%%%%%%%%%%%%%%%%%%%%%%%%%%%%%%%%%%%%%%%%%%%%%%%%%%%%%%%%%%%%%%%%%%%%%%%%%%%%%%%%%%%%%%%%%%%%%%%%%%%%%%%%%%%%%%%%%%%%%%%%%%%%%%%%%%%%%%%%%%%%%%%%%%%%%%%%%%%%%%%%%%%%%%%%%%%%%%%%%%%%%%%%%%%%%%%%%%%%%%%%%%%%%%%%%%%%%%%%%%%%%%%%%%%%%%%%%%%%%%%%%%%%%%%%%%%%%%%%%%%%%%%%%%
%%%%%%%%%%%%%%%%%%%%%%%%%%%%%%%%%%%%%%%%%%%%%%%%%%%%%%%%%%%%%%%%%%%%%%%%%%%%%%%%%%%%%%%%%%%%%%%%%%%%%%%%%%%%%%%%%%%%%%%%%%%%%%%%%%%%%%%%%%%%%%%%%%%%%%%%%%%%%%%%%%%%%%%%%%%%%%%%%%%%%%%%%%%%%%%%%%%%%%%%%%%%%%%%%%%%%%%%%%%%%%%%%%%%%%%%%%%%%%%%%%%%%%%%%%%%%%%%%%%%%%%%%%%%%%%%%%%%%%%%%%%%%%%%%%%%%%%%%%%%%%%%%%%%%%%%%%%%%%%%%%%%%%%%%%%%%%%%%%%%%%%%%%%%%%%%%%%%%%%%%%%%%%%%%%%%%%%%%%%%%%%%%%%%%%%%%%%%%%%%%%%%%%%%%%%%%%%%%%%%%%%%%%%%%%%%%%
% THE VERY END OF ENDS - ENDO OF ENDOS!
\end{document}